\documentclass{aa}

\usepackage{txfonts}
\usepackage{graphicx}
\usepackage{natbib}
\bibpunct{(}{)}{;}{a}{}{,}

\newcommand{\nwd}{5\,926}

\newcommand{\bparam}{0.24}
\newcommand{\bparame}{0.01}

\newcommand{\aparam}{0.14}
\newcommand{\aparame}{0.02}

\newcommand{\fhemin}{0.08}
\newcommand{\fhemine}{0.02}
\usepackage{color}

\begin{document}

\title{J-PLUS: Spectral evolution of white dwarfs by PDF analysis\thanks{The catalog with the atmospheric parameters and composition 
of the analyzed white dwarfs is available in electronic form 
both on the \texttt{jplus.WhiteDwarf} table at J-PLUS database and 
at the CDS via anonymous ftp to \url{cdsarc.u-strasbg.fr} (130.79.128.5)
or via \url{http://cdsarc.u-strasbg.fr/viz-bin/cat/J/A+A/658/A79}.}
}

\author{C.~L\'opez-Sanjuan\inst{\ref{CEFCA}}
\and P.~-E.~Tremblay\inst{\ref{warwick}}
\and A.~Ederoclite\inst{\ref{USP}}
\and H.~V\'azquez Rami\'o\inst{\ref{CEFCA}}
\and J.~M.~Carrasco\inst{\ref{IEEC}}
\and J.~Varela\inst{\ref{CEFCA}}
\and A.~J.~Cenarro\inst{\ref{CEFCA}}
\and A.~Mar\'{\i}n-Franch\inst{\ref{CEFCA}}
\and T.~Civera\inst{\ref{CEFCA}}
\and S.~Daflon\inst{\ref{ON}}
\and B.~T.~G\"ansicke\inst{\ref{warwick}}
\and N.~P.~Gentile~Fusillo\inst{\ref{ESO}}
\and F.~M.~Jim\'enez-Esteban\inst{\ref{INTA},\ref{VO}}
\and J.~Alcaniz\inst{\ref{ON}}
\and R.~E.~Angulo\inst{\ref{DIPC},\ref{ikerbasque}}
\and D.~Crist\'obal-Hornillos\inst{\ref{CEFCA}}
\and R.~A.~Dupke\inst{\ref{ON},\ref{MU},\ref{Alabama}}
\and C.~Hern\'andez-Monteagudo\inst{\ref{IAC},\ref{ULL}}
\and M.~Moles\inst{\ref{CEFCA}}
\and L.~Sodr\'e Jr.\inst{\ref{USP}}
}

\institute{Centro de Estudios de F\'{\i}sica del Cosmos de Arag\'on (CEFCA), Unidad Asociada al CSIC, Plaza San Juan 1, 44001 Teruel, Spain\\\email{clsj@cefca.es}\label{CEFCA}   
        \and
        Department of Physics, University of Warwick, Coventry, CV4 7AL, UK\label{warwick} 
        \and
        Instituto de Astronomia, Geof\'{\i}sica e Ci\^encias Atmosf\'ericas, Universidade de S\~ao Paulo, 05508-090 S\~ao Paulo, Brazil\label{USP}        
        \and
        Institut de Ci\`encies del Cosmos, Universitat de Barcelona (IEEC-UB), Mart\'{\i} i Franqu\`es 1, 08028 Barcelona, Spain\label{IEEC}
        \and
        Observat\'orio Nacional - MCTI (ON), Rua Gal. Jos\'e Cristino 77, S\~ao Crist\'ov\~ao, 20921-400 Rio de Janeiro, Brazil\label{ON}
        \and
        European Southern Observatory, Karl Schwarzschild Stra\ss e 2, Garching, 85748, Germany\label{ESO}
        \and
        Centro de Astrobiolog\'{\i}a (CSIC-INTA), ESAC Campus, Camino Bajo del Castillo s/n, 28692 Villanueva de la Ca\~nada, Spain\label{INTA}
        \and
        Spanish Virtual Observatory, 28692 Villanueva de la Ca\~nada, Spain\label{VO}
        \and
        Donostia International Physics Centre (DIPC), Paseo Manuel de Lardizabal 4, 20018 Donostia-San Sebastián, Spain\label{DIPC}
        \and
        IKERBASQUE, Basque Foundation for Science, 48013, Bilbao, Spain\label{ikerbasque}
        \and
        University of Michigan, Department of Astronomy, 1085 South University Ave., Ann Arbor, MI 48109, USA\label{MU}
        \and
        University of Alabama, Department of Physics and Astronomy, Gallalee Hall, Tuscaloosa, AL 35401, USA\label{Alabama}
        \and
        Instituto de Astrof\'{\i}sica de Canarias, La Laguna, 38205, Tenerife, Spain\label{IAC}
        \and
        Departamento de Astrof\'{\i}sica, Universidad de La Laguna, 38206, Tenerife, Spain\label{ULL}
}

\date{Received 8 July 2021 / Accepted 26 October 2021}

\abstract
{}
{We estimated the spectral evolution of white dwarfs with effective temperature using the Javalambre Photometric Local Universe Survey (J-PLUS) second data release (DR2), which provides 12 photometric optical passbands over $2\,176$ deg$^2$.}
{We analyzed $\nwd$ white dwarfs with $r \leq 19.5$ mag in common between a white dwarf catalog defined from {\it Gaia} EDR3 and J-PLUS DR2. We performed a Bayesian analysis by comparing the observed J-PLUS photometry with theoretical models of hydrogen- and helium-dominated atmospheres. We estimated the probability distribution functions for effective temperature ($T_{\rm eff}$), surface gravity, parallax, and composition; and the probability of having a H-dominated atmosphere ($p_{\rm H}$) for each source. We applied a prior in parallax, using {\it Gaia} EDR3 measurements as a reference, and derived a self-consistent prior for the atmospheric composition as a function of $T_{\rm eff}$.}
{We described the fraction of white dwarfs with a He-dominated atmosphere ($f_{\rm He}$) with a linear function of the effective temperature at $5\,000 < T_{\rm eff} < 30\,000$~K. We find $f_{\rm He} = \bparam \pm \bparame$ at $T_{\rm eff} = 10\,000$ K, a change rate along the cooling sequence of $\aparam \pm \aparame$ per 10~kK, and a minimum He-dominated fraction of $\fhemin \pm \fhemine$ at the high-temperature end. We tested the obtained $p_{\rm H}$ by comparison with spectroscopic classifications, finding that it is reliable. We estimated the mass distribution for the $351$ sources with distance $d < 100$~pc, mass $M > 0.45$~$M_{\odot}$, and $T_{\rm eff} > 6\,000$~K. The result for H-dominated white dwarfs agrees with previous studies, with a dominant $M = 0.59$~$M_{\odot}$ peak and the presence of an excess at $M \sim 0.8$~$M_{\odot}$. This high-mass excess is absent in the He-dominated distribution, which presents a single peak.}
{The J-PLUS optical data provide a reliable statistical classification of white dwarfs into H- and He-dominated atmospheres. We find a $21 \pm 3$\% increase in the fraction of He-dominated white dwarfs from $T_{\rm eff} = 20\,000$~K to $T_{\rm eff} = 5\,000$~K.}


\titlerunning{J-PLUS. White dwarf spectral evolution by PDF analysis}

\authorrunning{L\'opez-Sanjuan et al.}

\maketitle

\section{Introduction}\label{sec:intro}
White dwarfs are the degenerate remnants of stars with masses lower than $8-10$ $M_{\odot}$ and are the endpoint of stellar evolution for more than 97\% of Galactic stars \citep[e.g.,][and references therein]{ibeling13,doherty15}. White dwarfs are an important tool for studying the star formation history of the Milky Way, the late phases of stellar evolution, and can be used to improve our understanding of the physics of condensed matter.

Photometric white dwarf catalogs have so far mainly been based on the search for  objects showing an ultraviolet excess, such as the Palomar-Green catalog \citep[PG,][]{PGS}, the Kiso survey \citep[KUV,][]{KUV1,KUV2}, the Kitt Peak-Downes survey \citep[KPD,][]{KPD}, or the Sloan Digital Sky Survey \citep[SDSS,][]{sdss_dr16}; and using reduced proper motions (e.g., \citealt{NLTT,harris06,rowell11,GF15,munn17}). Subsequent spectroscopic follow-ups led to $\sim 35\,000$ white dwarfs with spectroscopic information \citep[e.g.,][]{eggen65,mccook99,eisenstein06,kepler19}. The main limitation of these photometric and spectroscopic catalogs is their nontrivial selection functions, a situation that has been improved thanks to the {\it Gaia} mission \citep{gaia}. {\it Gaia} provides a unique source of astrometric and photometric information that can be used to define the largest and most secure white dwarf catalog to date, with $360\,000$ sources so far \citep[][GF21 hereafter]{GF21}, and permits the definition of high-confidence volume-limited white dwarf samples \citep[e.g.,][]{hollands18,jimenezesteban18,gentilefusillo19,kilic20,mccleery20,gaia_edr3_nearby}.

Spectroscopic analysis of the early white dwarf catalogs demonstrated their spectral diversity \citep{sion83}. Those white dwarfs with hydrogen lines in their spectra are classified as DA type, while those presenting helium absorption can be DO (\ion{He}{II}) or DB (\ion{He}{I}). Featureless spectra in the optical define the continuum DC class, while the presence of heavy metals polluting the white dwarf atmosphere lead to the DZ and DQ (carbon) classifications. In addition to these general classes, hybrid types have also been reported (DAB, DBA, DZA, etc.). The dominant atmospheric composition of white dwarfs, defined as H-dominated or He-dominated depending on the most common component in their outer atmosphere, changes with cooling age (i.e., with decreasing effective temperature $T_{\rm eff}$). The DOs dominate at high temperatures ($T_{\rm eff} \gtrsim 80\,000$~K), with a steady decline in the fraction of He-dominated atmospheres down to $T_{\rm eff} \sim 40\,000$~K, where it reaches a minimum of 5\%-10\%. The fraction of He-dominated white dwarfs increases again at $T_{\rm eff} \sim 20\,000$~K towards lower temperatures, with a transition between DBs to DCs at $T_{\rm eff} \sim 11\,000$~K, where the \ion{He}{I} lines are no longer visible. This is also the case for H-dominated white dwarfs at $T_{\rm eff} \lesssim 5\,000$~K, where both H- and He-dominated white dwarfs are classified as DCs in the absence of spectral lines from polluting metals. This general picture is based on extensive observational work \citep{sion84,fleming86,greenstein86,fontaine87,eisenstein06db,tremblay08,giammichele12,limoges15,GB19,ourique19,blouin19,bedard20,cunningham20,mccleery20}.

The change in the fraction of He-dominated atmospheres in white dwarfs at $T_{\rm eff} \lesssim 25\,000$~K, referred to as spectral evolution hereafter, has been interpreted as the effect of convective dilution and convective mixing processes \citep[e.g.,][]{rolland18,cunningham20}, where the outer hydrogen layer and the underlying helium layer in white dwarfs are mixed because of the appearance of convection zones. In the dilution scenario, convection in the helium layer reaches the bottom of the thin hydrogen shell and erodes it, while in the mixing scenario the convection in the outermost hydrogen layer reaches the inner helium zone. In both cases, the net effect is to mix the hydrogen with the more abundant helium, producing the spectral change from a H-dominated to a He-dominated atmosphere for hydrogen mass below $\log M_{\rm H}/M \sim -6$. Moreover, the size of the convection zone increases with decreasing temperature, providing an indirect estimation of the mass of the hydrogen layer \citep[e.g.,][]{tremblay08,cunningham20} that can be compared with results from asteroseismology \citep{romero17}. Therefore, detailed knowledge of the  spectral evolution of white dwarfs provides clues about the physical processes acting in these objects as they cool over time.

Several of the above-mentioned studies of white dwarf spectral evolution at $T_{\rm eff} \lesssim 25\,000$~K are based on SDSS spectroscopy, which is affected by complicated selection effects \citep{GF15}. Unfortunately, current photometric data from the {\it Gaia} early data release three (EDR3) do not permit a spectral classification of the white dwarfs, preventing the study of their spectral evolution over the full sky. To mitigate this limitation, \citet{cunningham20} supplemented the {\it Gaia} catalog with optical and ultraviolet photometry, extending the white dwarf photometric classification down to $T_{\rm eff} = 9\,000$ K and obtaining a spectral evolution in agreement with the spectroscopic results.

In the present study, we used the 12 optical bands of the Javalambre Photometric Local Universe Survey (J-PLUS; \citealt{cenarro19}) second data release (DR2; \citealt{jplus_dr2}) over $2\,176$ deg$^2$ to supplement the GF21 catalog, which is based on {\it Gaia} EDR3, and to provide observational constraints to the white dwarf spectral evolution in the range $5\,000 < T_{\rm eff} < 30\,000$~K. The J-PLUS photometric system (Table~\ref{tab:JPLUS_filters}) is composed of five SDSS-like ($ugriz$) and seven medium-band filters located in key stellar features, such as the $4\,000~\AA$ break ($J0378$, $J0395$, $J0410$, and $J0430$), the Mg $b$ triplet ($J0515$), H$\alpha$ at rest frame ($J0660$), and the calcium triplet ($J0861$). We used this extra photometric information, coupled with a Bayesian analysis of the data, to disentangle the white dwarf spectral type over a wide range of effective temperatures.

This paper is organized as follows. In Sect.~\ref{sec:data} we detail the J-PLUS photometric data and the reference white dwarf catalog from {\it Gaia} EDR3 used in our analysis. The Bayesian fitting process is described in Sect.~\ref{sec:methods}. The final selection of the sample and the derived atmospheric parameters are presented in Sect.~\ref{sec:teff_logg}. The white dwarf spectral evolution from J-PLUS is reported in Sect.~\ref{sec:fnonda_teff}. Finally, a summary and the conclusions of our work are presented in Sect.~\ref{sec:conclusions}. All magnitudes are expressed in the AB system \citep{oke83}.


\begin{table} 
\caption{J-PLUS passbands, including filter transmission, CCD efficiency, telescope optics, and atmosphere.}
\label{tab:JPLUS_filters}
\centering 
        \begin{tabular}{l c c}
        \hline\hline\rule{0pt}{3ex} 
        Passband   & Effective wavelength & Rectangular width  \\
                   &   [nm]             &   [nm]\\
        \hline\rule{0pt}{2ex}
        $u$             &353.6  &  34.3     \\ 
        $J0378$         &378.2  &  13.8     \\ 
        $J0395$         &393.9  &   9.9     \\ 
        $J0410$         &410.8  &  19.4     \\ 
        $J0430$         &430.3  &  19.6     \\ 
        $g$             &481.0  & 129.5     \\ 
        $J0515$         &514.1  &  20.5     \\ 
        $r$             &627.2  & 143.4     \\ 
        $J0660$         &660.4  &  14.6     \\ 
        $i$             &766.9  & 139.7     \\ 
        $J0861$         &861.1  &  40.2     \\ 
        $z$             &898.0  & 124.9     \\ 
        \hline 
\end{tabular}
\end{table}

\begin{figure*}[t]
\centering
\resizebox{0.49\hsize}{!}{\includegraphics{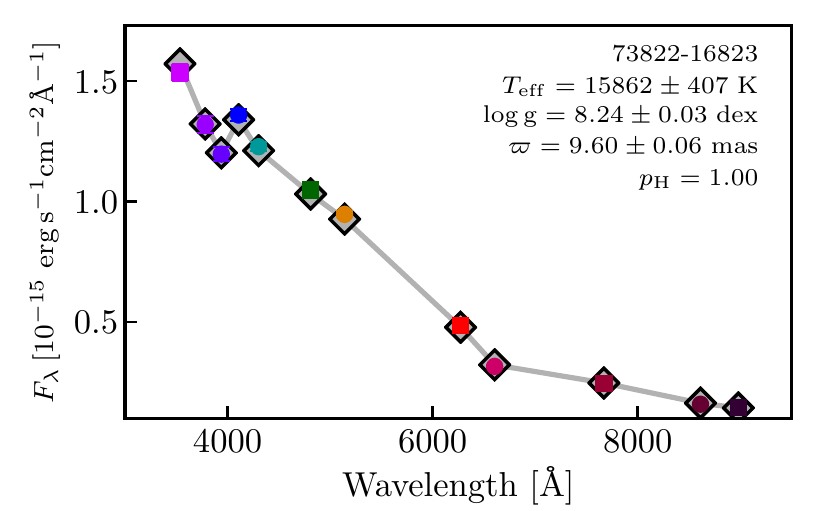}}
\resizebox{0.49\hsize}{!}{\includegraphics{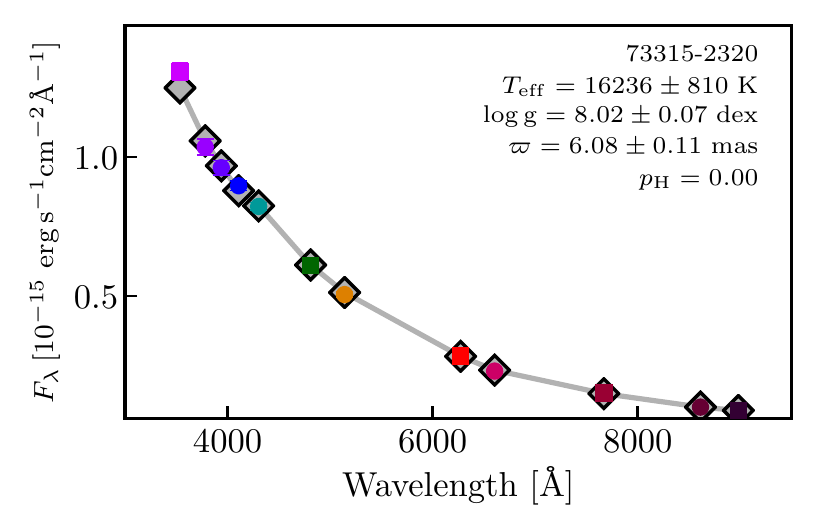}}
\resizebox{0.49\hsize}{!}{\includegraphics{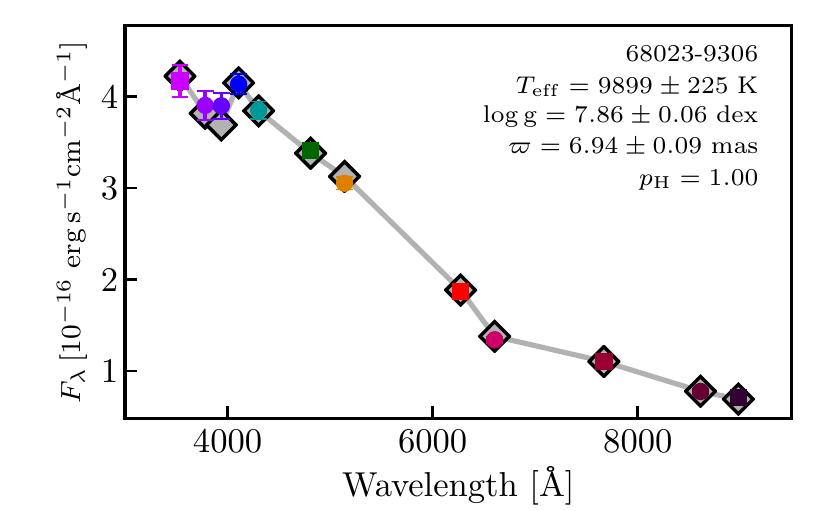}}
\resizebox{0.49\hsize}{!}{\includegraphics{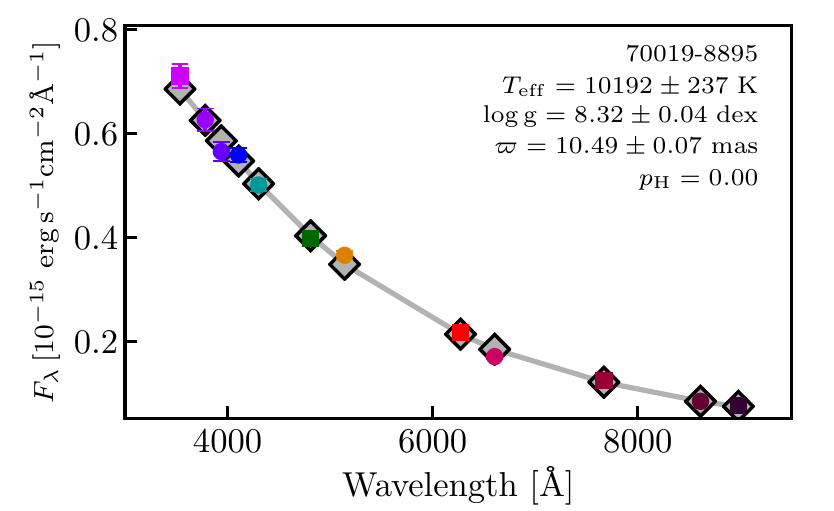}}
\caption{Spectral energy distributions of four white dwarfs analyzed as part of this work, selected to highlight the difference between H- and He-dominated atmospheres in the J-PLUS passbands. The colored points in all the panels are the 3 arcsec diameter photometry corrected to total magnitudes from J-PLUS (squares for broad bands, $ugriz$; circles for medium bands, $J0378$, $J0395$, $J0410$, $J0430$, $J0515$, $J0660$, and $J0861$). The gray diamonds connected with a solid line show the best-fitting solution, with the derived parameters and their uncertainties labeled in the panels: effective temperature ($T_{\rm eff}$), surface gravity ($\log {\rm g}$), parallax ($\varpi$), and probability of having a H-dominated atmosphere ($p_{\rm H}$). The unique J-PLUS identification, composed by the \texttt{TILE\_ID} of the reference $r-$band image and the \texttt{NUMBER} assigned by SExtractor to the source, is also reported in the panels for reference. Sources with $T_{\rm eff} \sim 16\,000$ K ({\it top panels}) and $T_{\rm eff} \sim 10\,000$ K ({\it bottom panels}) are presented. {\it Left panels}: Sources with H-dominated atmospheres, $p_{\rm H} = 1$. {\it Right panels}: Sources with He-dominated atmospheres, $p_{\rm He} = 0$.}
\label{fig:examples}
\end{figure*}

\section{Data}\label{sec:data}

\subsection{J-PLUS photometric data}\label{sec:jplus}
J-PLUS\footnote{\url{www.j-plus.es}} is being conducted from the Observatorio Astrof\'{\i}sico de Javalambre (OAJ, Teruel, Spain; \citealt{oaj}) using the 83\,cm Javalambre Auxiliary Survey Telescope (JAST80) and T80Cam, a panoramic camera of 9.2k $\times$ 9.2k pixels that provides a $2\deg^2$ field of view (FoV) with a pixel scale of 0.55 arsec pix$^{-1}$ \citep{t80cam}. The J-PLUS filter system is composed of 12 passbands (Table~\ref{tab:JPLUS_filters}). The J-PLUS observational strategy, image reduction, and main scientific goals are presented in \citet{cenarro19}.

The J-PLUS DR2 comprises $1\,088$ pointings ($2\,176$ deg$^2$) observed, reduced, and calibrated in all survey bands \citep{jplus_dr2, clsj21zsl}. The limiting magnitudes (5$\sigma$, 3 arsec aperture) of the DR2 are $\sim 22$ mag in $g$ and $r$ passbands, and $\sim 21$ mag in the another ten bands. The median point spread function (PSF) full width at half maximum (FWHM) in the DR2 $r$-band images is 1.1 arcsec. Source detection was done in the $r$ band using \texttt{SExtractor} \citep{sextractor}, and the flux measurement in the 12 J-PLUS bands was performed at the position of the detected sources using the aperture defined in the $r$-band image. Objects near the borders of the images, close to bright stars, or affected by optical artefacts were masked from the initial $2\,176$ deg$^2$, providing a high-quality area of $1\,941$ deg$^2$. The DR2 is publicly available on the J-PLUS web site\footnote{\url{www.j-plus.es/datareleases/data_release_dr2}}.

We used aperture photometry of  3 arcsec in diameter to analyze the white dwarf population. The observed fluxes were stored in the vector $\vec{f} = \{ f_j \}$, and their errors were stored in the vector $\sigma_{\vec{f}} = \{\sigma_j\}$, where the index $j$ runs the J-PLUS passbands. The error vector includes the uncertainties from photon counting, sky background, and photometric calibration \citep{clsj21zsl}.

\subsection{{\it Gaia} white dwarf catalog}\label{sec:gf21}
We used the {\it Gaia}-based catalog of white dwarfs presented in GF21 as a reference, and we did not attempt to derive a white dwarf catalog from J-PLUS photometry. The main reason for this choice is that the parallax information from {\it Gaia} EDR3 permits the estimation of the white dwarf surface gravity, which is poorly constrained from photometry alone, and the definition of volume-limited samples. In combination with the 12-band J-PLUS photometry in the present work, the atmospheric parameters of the common white dwarfs, including the atmospheric composition, can be computed (Sect.~\ref{sec:methods}). As a drawback, the selection effects of the GF21 catalog will be inherited by our common sample.

As a summary of the selection process performed by GF21, $1\,280\,266$ objects are selected using several quality flags and their location in the Hertzsprung-Russell diagram of {\it Gaia} EDR3. This initial selection represents a compromise between removing the majority of sources with non-optimal {\it Gaia} measurements and preserving all the stars in the white dwarf locus.

A total of $22\,998$ spectroscopically confirmed single white dwarfs and $7\,124$ contaminant objects obtained from SDSS DR16 \citep{sdss_dr16} were then used to map their distribution in the absolute $G$-band magnitude versus color space and to assign a white dwarf probability, $P_{\rm WD}$. Following the prescriptions in GF21, we selected the $259\,073$ sources with $P_{\rm WD} > 0.75$, of which $25\,632$ have SDSS spectroscopy. This spectroscopic sample comprises $91$\% confirmed white dwarfs, $1$\% contaminant objects, $3$\%  white dwarf--main sequence binaries or cataclysmic variables, and the rest have unreliable classification. When comparing with confirmed SDSS spectroscopic white dwarfs, GF21 also find no significant color bias in the selection. We refer the reader to GF21 for a detailed description of the selection criteria and the properties of the reference sample.

We cross-matched the $259\,073$ sources with $P_{\rm WD} > 0.75$ in the GF21 catalog with the J-PLUS DR2 dataset using a $1.5$ arcsec radius, finding $11\,182$ sources with $r \leq 20.3$ mag in common. The final white dwarf sample, which ensures a well-defined volume and magnitude selection, is detailed in Sect.~\ref{sec:teff_logg}.


\section{Estimation of white dwarf atmospheric parameters and composition}\label{sec:methods}
We aim to estimate the following probability density function (PDF) for each white dwarf in the sample,
\begin{equation}
    {\rm PDF}\,(t,\theta\,|\,\vec{f}, \vec{\sigma}_{\vec{f}}) \propto \mathcal{L}\,(\,\vec{f}\,|\,t,\theta,\sigma_{\vec{f}}) \times P\,(\theta) \times P\,(t),
\end{equation}
where $\theta$ are the parameters in the fitting, $t$ are the different atmospheric compositions considered in our analysis, $\mathcal{L}$ is the likelihood of the data for a given set of parameters and composition, and $P$ are the prior probabilities. The final PDF is normalized to one by definition:
\begin{equation}
    \sum_t \int {\rm PDF}\,(t,\theta)\,{\rm d}\theta = 1.
\end{equation}

The parameters in the fitting were $\theta = \{T_{\rm eff}, \log {\rm g}, \varpi \}$, corresponding to effective temperature, surface gravity, and parallax. We explored two atmospheric compositions, corresponding to hydrogen- and helium-dominated atmospheres. We assumed that H-dominated atmospheres correspond to DA spectral type, while He-dominated atmospheres to DB and DC spectral types. The latter assumption imposes a lower limit on effective temperature  in our analysis of $T_{\rm eff} = 5\,000$ K, because at lower temperatures the hydrogen lines are no longer visible in H-dominated atmospheres and they are also classified as DCs \citep[e.g.,][]{greenstein88}. For simplicity in the notation, hybrid spectral types such as DABs and DBAs were considered as their main composition type. We therefore used $t = \{ {\rm H}, {\rm He} \}$ in our analysis.

The probability of being H-dominated is therefore defined as
\begin{equation}
    p_{\rm H} = \int {\rm PDF}\,({\rm H},\theta)\,{\rm d}\theta,
\end{equation}
and the probability of being He-dominated is $p_{\rm He} = 1 - p_{\rm H}$.

In the following sections, we describe the likelihood and the priors used in the analysis of J-PLUS + {\it Gaia} white dwarfs. We present four representative examples in Fig.~\ref{fig:examples}.

\subsection{Likelihood}\label{sec:likelihood}
We defined the likelihood of the data given a set of parameters as
\begin{equation}
    \mathcal{L}\,(\,\vec{f}\,|\,t,\theta,{\boldmath \sigma}_{\vec{f}}) = \prod_{j = 1}^{12} P_{\rm G}\,(f_j\,|\,f_{j}^{\rm mod}, \sigma_{j}),
\end{equation}
where the index $j$ runs over the 12 J-PLUS passbands, the function $P_{\rm G}$ defines a Gaussian probability distribution, 
\begin{equation}
P_{\rm G}\,(x\,|\,\mu,\sigma) = \frac{1}{\sqrt{2\pi}\,\sigma}\ {\rm exp}\Big[-\frac{(x - \mu)^2}{2\sigma^2}\Big] = \frac{1}{\sqrt{2\pi}\,\sigma}\ {\rm exp}\Big[\frac{- \chi^2}{2}\Big],
\end{equation}
and the model flux was estimated as
\begin{equation}
    f_{j}^{\rm mod}\,(t,\theta)= \bigg( \frac{\varpi}{100} \bigg)^2\,F_{t,j}\,(T_{\rm eff},\log {\rm g})\,10^{-0.4\,k_j\,E(B-V)}\,10^{0.4\,C^{\rm aper}_j},
\end{equation}
where $E(B-V)$ is the assumed color excess of the white dwarf, $k_j$ is the extinction coefficient, $C^{\rm aper}_j$ is the aperture correction needed to translate the observed $3$ arcsec magnitudes to total magnitudes, and $F_{t,j}$ is the theoretical absolute flux emitted by a white dwarf of type $t$ located at 10 pc distance, that is, with a parallax of $100$ mas.

We assumed pure-H models to describe H-dominated atmospheres ($t = {\rm H}$, \citealt{tremblay11,tremblay13}), while mixed models with H/He = $10^{-5}$ at $T_{\rm eff} > 6\,500$ K and pure-He models at $T_{\rm eff} < 6\,500$ K were used to describe He-dominated atmospheres ($t = {\rm He}$, \citealt{cukanovaite18, cukanovaite19}). The mass--radius relation of \citet{fontaine01} for thick (H-atmospheres) and thin (He-atmospheres) hydrogen layers was assumed in the modeling. An extensive discussion about these choices and extra details of the models are presented in \citet{bergeron19}, \citet{GF20}, \citet{mccleery20}, and GF21.

The likelihood was estimated in a dense grid of models with $\Delta \log T_{\rm eff} = 0.005$ dex, $\Delta \log {\rm g} = 0.005$ dex, and $\Delta \varpi = 0.05$ mas. The fluxes for those parameters not included in the initial grid of theoretical models were estimated by linear interpolation.

The color excess was estimated using the $E(B-V)$ at infinity from \citet{sfd98}, properly scaled at a distance of $d = 1/\varpi$ with the Milky Way dust model presented in \citet{li18}. The uncertainty in $E(B-V)$ was fixed to $0.012$ mag. This error was estimated from the dispersion in the comparison between the color excess directly measured from the star-pair method \citep{yuan13} with the assumed $E(B-V)$. We note that this extinction scheme was used in the photometric calibration of J-PLUS DR2 \citep{clsj21zsl}, so we decided to also follow it here for consistency. The extinction coefficients for J-PLUS passbands are also reported in \citet{clsj21zsl}.

The aperture correction was defined as
\begin{equation}
    C^{\rm aper} = C^{\rm tot}_{6} + C^{6}_{3},
\end{equation}
where the first term is the correction from $6$ arcsec magnitudes to total magnitudes, and the second term is the needed correction to go from $3$ arcsec to $6$ arcsec photometry. The 6 arcsec to total correction was estimated from the growth curve of bright, nonsaturated stars for each passband and pointing. For each star, increasingly large circular apertures were measured until convergence within errors. This defined the aperture size that provides the total magnitude of the sources in the pointing that is then compared with the magnitude at $6$ arcsec aperture to provide $C_{6}^{\rm tot}$.

The signal-to-noise ratio (S/N) from 3 arcsec photometry ($m_{3}$) is 30\% to 50\% larger than from 6 arcsec photometry ($m_{6}$) at $r = 19.5$ mag, the limiting magnitude of the present study; however, it is affected by PSF variation among passbands and along the FoV. To overcome these limitations, we estimated the 3 arcsec to 6 arcsec aperture correction using the measurements in the J-PLUS catalog. For each tile and passband, we derived the $c_{63}= m_{6} - m_{3}$ magnitude for objects with a stellarity parameter\footnote{We used the variable \texttt{sglc\_prob\_star} in the J-PLUS DR2 database as morphological classifier \citep{clsj19psmor}} larger than 0.9. We used the 250 brightest stars in the tile to estimate the median $c_{63}$ in a $5 \times 5$ grid in $(X,Y)$ to enhance the signal and ensure a smooth variation along the FoV. We parameterized the $(X,Y)$ variation with a combination of Chebysev polynomial up to second order. The resulting smooth function in $(X,Y)$ provides the aperture correction $C^{\rm 6}_{3}$ to transform 3 arcsec photometry to 6 arcsec photometry. The uncertainty in this correction, estimated from the dispersion of the $c_{63}$ measurements with respect to the best-fitting model after accounting for the observational errors, is typically below $5$ mmag. Because of its limited impact in the final error budget, the total aperture correction $C^{\rm aper}$ was assumed with no uncertainty in the fitting process.

We applied both the extinction and the aperture correction to the model fluxes instead of attempting to correct the observations. This is motivated by the fact that application of such factors to negative fluxes produces ill-defined values. The attenuation of model fluxes is always positive and well-defined, and permits the proper statistical treatment of those observations close to the detection limit of the survey.

\subsection{Priors}
The application of priors in the estimation of white dwarf parameters helps to break degeneracies and to avoid nonphysical or unlikely solutions \citep[e.g.,][]{mortlock09,omalley13}. We applied a prior in the parallax and the atmospheric composition, as detailed in the following sections.

\subsubsection{Parallax prior}\label{sec:prior_px}
The parallax prior was
\begin{equation}
    P\,(\varpi) = P_{\rm G}\,(\varpi\,|\,\varpi_{\rm EDR3}, \sigma_{\varpi}),
\end{equation}
where $\varpi_{\rm  EDR3}$ and $\sigma_{\varpi}$ are the parallax and its error obtained from ${\it Gaia}$ EDR3 \citep{gaiaedr3,lindegren21a}. The published values of the parallax were corrected by the {\it Gaia} zero-point offset following the prescription by \citet{lindegren21b}, already validated by several independent studies to a few $\mu$as level \citep[e.g.,][]{huang21par,ren21par,maiz21par}.

We note that the assumption of a mass--radius relation in the estimation of the theoretical fluxes coupled the parallax and the surface gravity variables. In this framework, the prior in $\varpi$ also imposes strong conditions on $\log {\rm g}$, which is poorly constrained by photometry alone. This issue is explored in Sect.~\ref{sec:chidist}.

\subsubsection{Atmospheric composition prior}\label{sec:prior_da}
The used prior in the atmospheric composition was
\begin{equation}
    P\,({\rm H}\,|\,T_{\rm eff}) = 1 - f_{\rm He}\,(T_{\rm eff}),\label{eq:fdbprior}
\end{equation}
where $f_{\rm He}$ is the fraction of helium-dominated white dwarfs,
\begin{equation}
   f_{\rm He} = \frac{N_{\rm He}}{N_{\rm H} + N_{\rm He}},
\end{equation}
$N_{\rm H}$ is the number of H-dominated white dwarfs, and $N_{\rm He}$ is the number of He-dominated white dwarfs at a given effective temperature. We note that this prior distribution is equivalent to the white dwarf spectral evolution with $T_{\rm eff}$, the elucidation of which is the main goal of our work.

Here, we present the technical details in the computation of $f_{\rm He}$, while the obtained results are detailed in Sect.~\ref{sec:fnonda_teff}. We can obtain the fraction of He-dominated white dwarfs from the final PDFs of the population as
\begin{equation}
    f_{\rm He}\,(T_{\rm eff}) = \frac{\sum_i V^{-1}_i\,({\rm He})\times{\rm PDF}_i\,({\rm He}, T_{\rm eff})}{\sum_i \sum_t V_i^{-1}\,(t) \times {\rm PDF}_i\,(t,T_{\rm eff})},\label{eq:fdbpdf}
\end{equation}
where the index $i$ refers to the white dwarfs in the sample, the full PDFs were marginalized over the nonexplicit variables, and $V\,(t)$ is the maximum effective volume probed by each white dwarf as computed in Sect.~\ref{sec:selec}. This defines a continuum variable in effective temperature and a histogram can be created by integrating over $T_{\rm eff}$ bins. In the latter case, the errors in $f_{\rm He}$ were estimated by bootstrapping.

We note that $f_{\rm He}\,(T_{\rm eff})$ is present both in Eq.~(\ref{eq:fdbprior}) and Eq.~(\ref{eq:fdbpdf}). In other words, the input prior used to estimate the PDFs should be similar to the output fraction derived from the posteriors. Instead of imposing an externally computed prior, which would compromise the interpretation and significance of our results, we used the above argument to derive a prior from J-PLUS data in a self-consistent way and to estimate $f_{\rm He}\,(T_{\rm eff})$. Formally, this is a Bayesian hierarchical analysis where the parameters of the prior function are derived from the data. The assumed function in $T_{\rm eff}$ for the He-dominated fraction and the parameters that describe the white dwarf spectral evolution are presented and justified in Sect.~\ref{sec:fnonda_teff}.

To compute the self-consistent prior, we estimated the aggregated $\chi^2$ between the output PDF-based histogram and the binned input prior using the same $T_{\rm eff}$ ranges in both cases. We included the covariance terms between different temperature bins in the process, which were estimated from bootstrapping. We minimized the $\chi^2$ by exploring the parameters that define $f_{\rm He}\,(T_{\rm eff})$ with the \texttt{emcee} code \citep{emcee}, a \texttt{Python} implementation of the affine-invariant ensemble sampler for the Markov chain Monte Carlo (MCMC) technique proposed by \citet{goodman10}. The \texttt{emcee} code provides a collection of solutions in the parameter space, with the density of solutions being proportional to the posterior probability of the parameters. We obtained the central values of the parameters and their uncertainties from a Gaussian fit to the distribution of the solutions.

\subsection{Selection probability and derived quantities}\label{sec:selec}
In addition to the fundamental parameters obtained in the fitting process, other properties of the analyzed white dwarfs can be derived, such as the mass or the luminosity.

The mass PDF for a given type $t$ was estimated as
\begin{equation}
    {\rm PDF}\,(M\,|\,t) = \int {\rm PDF}\,(t,\theta) \times \delta[M - \mathcal{M}\,(t,\theta)]\,{\rm d}\theta,
\end{equation}
where $\delta$ is the Dirac delta function, and $\mathcal{M}$ is the mass predicted for each model. The same procedure was used to obtain the PDF in $\hat{r}$, the total de-reddened $r$-band apparent magnitude, as
\begin{equation}
    {\rm PDF}\,(\hat{r}\,|\,t) = \int {\rm PDF}\,(t,\theta) \times \delta[\hat{r} - \mathcal{R}\,(t,\theta)]\,{\rm d}\theta,
\end{equation}
where
\begin{equation}
    \mathcal{R}\,(t,\theta) = -2.5\log_{10} [ f_r\,(t,\theta) ] + 46.8,\end{equation}
and
\begin{equation}
        f_r\,(t,\theta) = \bigg( \frac{\varpi}{100} \bigg)^2\,F_{t,r}\,(T_{\rm eff},\log {\rm g}).
\end{equation}
We note that the variable $\hat{r}$ can be used to define a well-controlled magnitude-limited sample. We defined the selection probability as
\begin{equation}
    p_{\rm sel}\,(r_{\lim}) = \sum_t \int_{-\infty}^{r_{\rm lim}}\,{\rm PDF}\,(\hat{r}\,|\,t)\,{\rm d}\hat{r},
\end{equation}
where the limiting magnitude $r_{\rm lim}$ defines the selection of the sample. In addition, we imposed a minimum of 10 pc and a maximum of $1$ kpc in the posterior analysis. In other words, the estimation of the PDF was performed in the full distance range, but we only kept those solutions with parallax $\varpi \in [1,100]$ mas. This selection helps to avoid extreme solutions and provides well-defined volumes. The final selection of the white dwarf sample and its distribution in $\hat{r}$ are presented in Sect.~\ref{sec:counts}.

Finally, we estimated the effective volume probed by each white dwarf in the sample. This quantity is needed to account for the different luminosities (i.e., probed volumes) of H- and He-dominated white dwarfs for a given effective temperature and surface gravity. We defined the effective volume as
\begin{equation}
    V\,(t) = \frac{\int {\rm PDF}\,(t,\theta) \times \mathcal{V}\,(t,\theta)\,{\rm d}\theta}{\int {\rm PDF}\,(t,\theta)\,{\rm d}\theta}\ \ \ \ \rm{[kpc^{3}]},
\end{equation}
where
\begin{equation}
    \mathcal{V}\,(t,\theta) = \frac{4}{3}\pi f_{\Omega}\,(\varpi_{\rm min}^{-3} - \varpi_{\rm max}^{-3})\ \ \ \ \rm{[kpc^{3}]},
\end{equation}
 the fraction of the sphere subtended by the unmasked J-PLUS DR2 is $f_{\Omega} = 0.047$ ($1\,941$ deg$^2$), the maximum parallax was set to $\varpi_{\rm max} = 100$ mas, the minimum parallax was defined as
\begin{equation}
    \varpi_{\rm min}\,(t,\theta) = {\rm max}[1, \varpi_{\rm lim}],
\end{equation}
and $\varpi_{\rm lim}$ is the parallax associated to the $r_{\rm lim}$ magnitude for a given set of parameters (i.e, the maximum distance at which a white dwarf can be observed given the selection magnitude $r_{\rm lim}$). The effective volume is used in the estimation of the spectral evolution of  white dwarfs, as detailed in Sect.~\ref{sec:prior_da}.

\subsection{Summary statistics}
We obtained the summary statistics for each parameter by marginalizing the PDFs over the other parameters at a fixed atmospheric composition and performing a Gaussian fit to the resulting distribution. The retrieved parameters and their uncertainties are the median and the dispersion of the best-fitting Gaussian. We find that this process provides a proper description of the posteriors and allows us to gather the relevant information table format.

The reported values of $\hat{r}$ were estimated before the final selection in both magnitude and distance (Sect.~\ref{sec:selec}), while the remaining variables were computed from the remaining solutions after applying the selection to the posterior PDFs. The summary statistics are publicly available in the J-PLUS database and the Centre de Donn\'ees astronomiques de Strasbourg\footnote{\url{https://cds.u-strasbg.fr}} (CDS). A description of and links to the data are presented in Appendix~\ref{app:data}.

We also provide $p_{\rm H}$ in the table. In those cases where the probability of having a H-dominated atmosphere is different from both zero and one, we included the atmospheric parameters for both H- and He-dominated atmospheres. These values should be weighted with the corresponding $p_{\rm H}$ to provide meaningful results, as illustrated in Sect.~\ref{sec:massdist}.


\section{White dwarf sample and atmospheric parameters}\label{sec:teff_logg}
We present in this section the atmospheric parameters obtained in the analysis of the J-PLUS DR2 + ${\it Gaia}$ EDR3 catalog. The final selection of the sample is described in Sect.~\ref{sec:counts} and we analyze the quality of the fitting process in Sect.~\ref{sec:chidist}. The obtained effective temperatures and surface gravities are studied in Sect.~\ref{sec:tlogg}.

\begin{figure}[t]
\centering
\resizebox{\hsize}{!}{\includegraphics{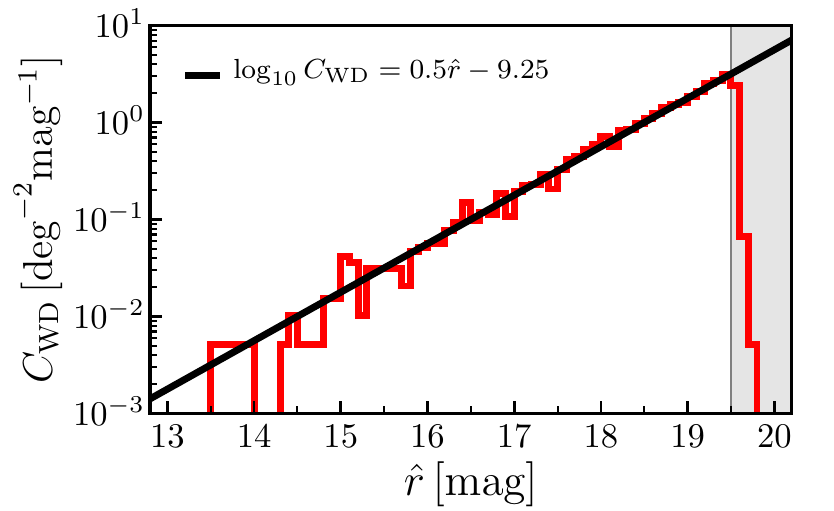}}
\caption{White dwarf number counts in J-PLUS DR2 as a function of the total and de-reddened $r$-band apparent magnitude, noted $\hat{r}$ (red histogram). The black line is the function that better describes the distribution, as labeled in the panel. The white area shows the magnitudes used to define the white dwarf sample, $\hat{r} \leq 19.5$ mag.}
\label{fig:rdist}
\end{figure}

\begin{figure}[t]
\centering
\resizebox{\hsize}{!}{\includegraphics{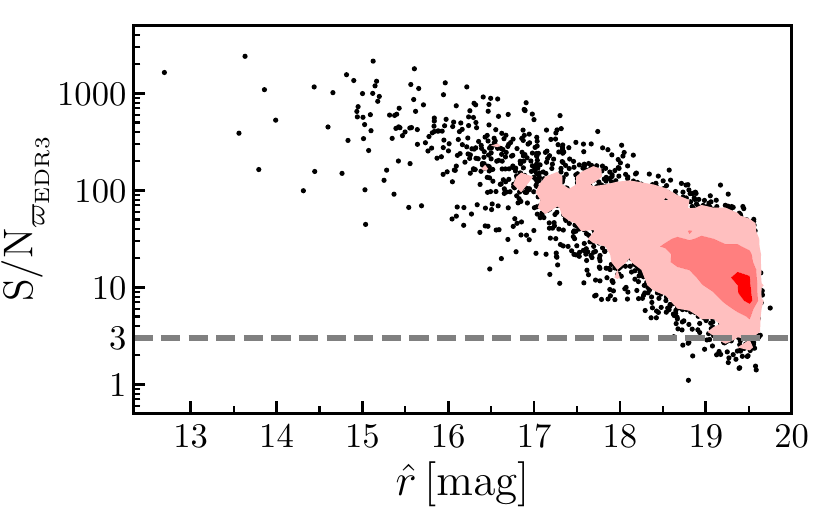}}
\caption{Signal-to-noise ratio in the {\it Gaia} EDR3 parallax (${\rm S/N}_{\varpi_{\rm EDR3}}$) as a function of the de-reddened $r$-band apparent magnitude ($\hat{r}$) for the final white dwarf sample. The black dots show individual measurements. The red areas from lighter to darker enclose 90\%, 50\%, and 10\% of the sources, respectively. The gray dashed line marks ${\rm S/N}_{\varpi_{\rm EDR3}} = 3$.}
\label{fig:snrpx}
\end{figure}

\subsection{Final selection and number density}\label{sec:counts}
We estimated the atmospheric parameters and composition for the $11\,182$ sources in common between the GF21 catalog and J-PLUS DR2 at $r < 20.3$ mag (Sect.~\ref{sec:gf21}), and only kept those with a selection probability $p_{\rm sel} > 0.01$ for $\varpi \in [1,100]$ mas and $\hat{r} \leq 19.5$ mag. We refer the reader to Sect.~\ref{sec:selec} for further details about the definition of the selection probability. This selection ensures enough S/N in the J-PLUS photometry to perform a meaningful statistical analysis of the sources and provides well-defined volumes.

The final sample comprises $\nwd$ white dwarfs, which translates to a number density of $2.8$ deg$^{-2}$. We present the white dwarf number counts in Fig.~\ref{fig:rdist}, obtained as the $p_{\rm sel}$-weighted histogram in the dust de-reddened $\hat{r}$ magnitude normalized by the J-PLUS DR2 surveyed area and the magnitude bin size. We note that the used $\hat{r}$ magnitudes refer to the full range of solutions, that is, before applying the selection constraints. Hence, sources with $\hat{r} > 19.5$ and $p_{\rm sel} < 1$ are present in the sample. The number counts are well described as
\begin{equation}
    \log C_{\rm WD} = 0.5\,\hat{r} - 9.25\ \ \ \ {\rm [deg^{-2} mag^{-1}]}.
\end{equation}
We find no evidence of departure from linearity in log scale over the six magnitudes covered by our study, $13.5 \lesssim \hat{r} \leq 19.5$ mag.
We present the S/N in the {\it Gaia} EDR3 parallax as a function of $\hat{r}$ for the final sample in Fig.~\ref{fig:snrpx}. As expected, there is a trend towards lower signal at fainter magnitudes. The median value for the sample is ${\rm S/N} = 20.3$, with more than 99\% of the sources having ${\rm S/N} > 3$.

Finally, we estimated the purity of our final sample using the $42\,007$ spectral classifications based on SDSS DR16 \citep{sdss_dr16} also presented in the GF21 catalog. We cross-matched the spectroscopic catalog with our final sample, discarding duplicate entries and sources with unreliable classification. This provided a total of $1\,835$ sources with spectral classification, with $1\,805$ ($98.3$\%) white dwarfs, $25$ ($1.4$\%) white dwarf--main sequence binaries or cataclysmic variables, and $5$ ($0.3$\%) contaminants or objects with unknown classification. We conclude that our final sample of $5\,926$ white dwarfs presents a well-defined magnitude and volume selection with a purity above $98$\%.

\begin{figure}[t]
\centering
\resizebox{\hsize}{!}{\includegraphics{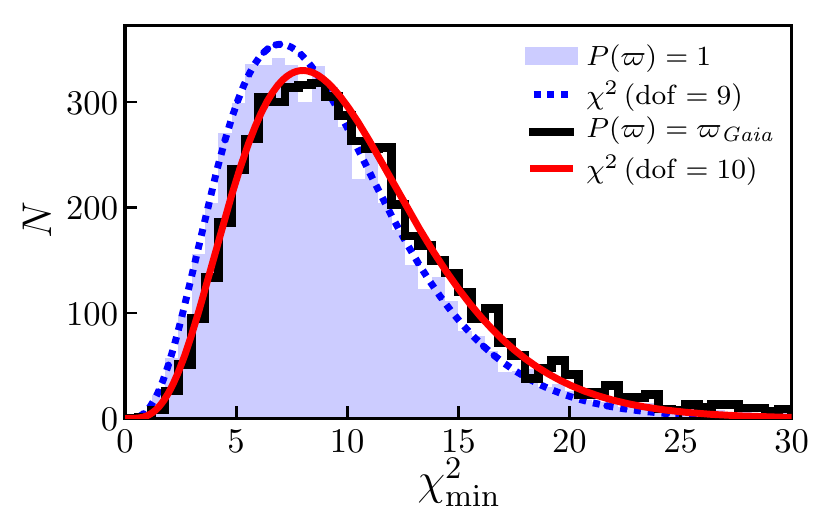}}
\caption{Distribution of $\chi^2_{\rm min}$ for the white dwarf sample. The open black histogram represents the solutions when the ${\it Gaia}$ EDR3 parallax was used as prior, and the blue solid histogram shows when no prior in parallax was used in the fitting. The red solid and blue dotted lines mark the expected $\chi^2$ distribution for 10 and 9 degrees of freedom, respectively.}
\label{fig:chi2dist}
\end{figure}

\begin{figure}[t]
\centering
\resizebox{\hsize}{!}{\includegraphics{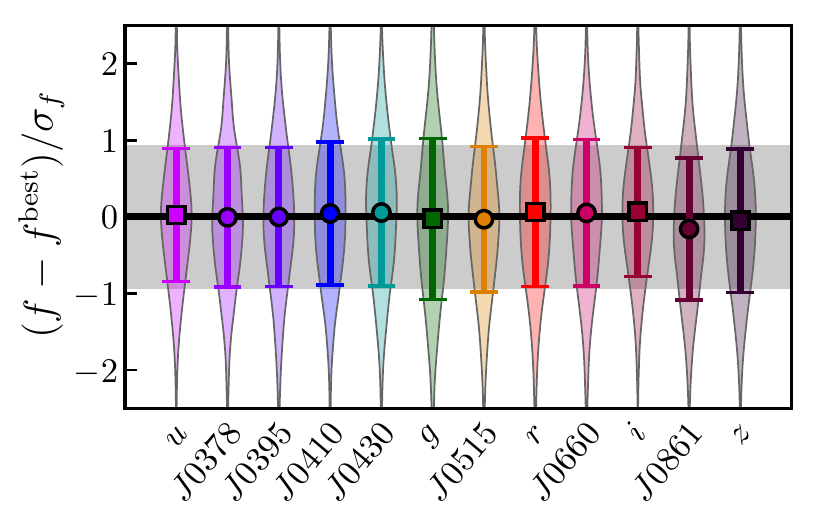}}\caption{Distribution of the observed flux minus the best-fitting flux normalized by the photometric error. The squares (broad bands) and circles (medium bands) show the median and the dispersion of the distribution, depicted with the violin plots. The black solid line marks a zero difference, and the gray area shows the $\pm 0.91$ value expected for the dispersion.}
\label{fig:chidist}
\end{figure}

\begin{figure*}[t]
\centering
\resizebox{0.49\hsize}{!}{\includegraphics{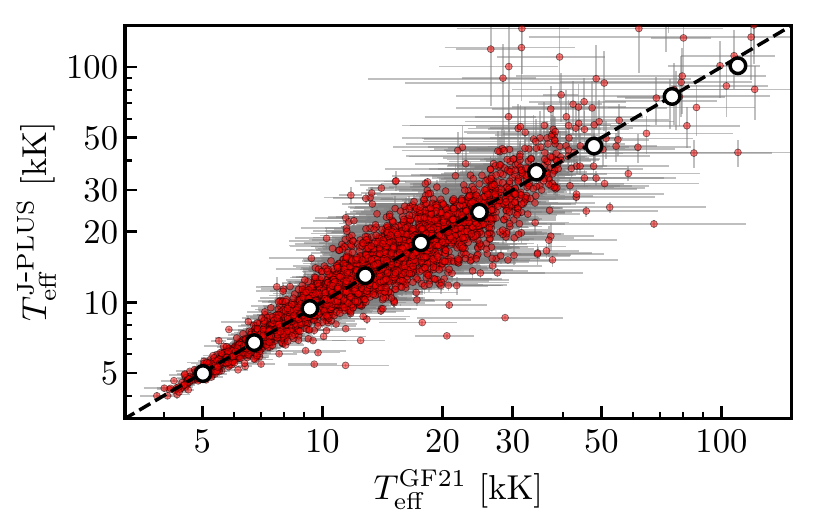}}
\resizebox{0.49\hsize}{!}{\includegraphics{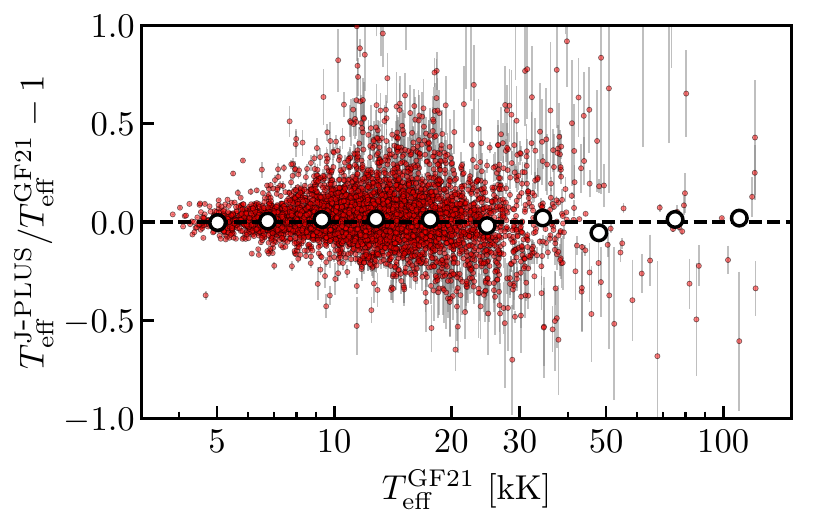}}

\resizebox{0.49\hsize}{!}{\includegraphics{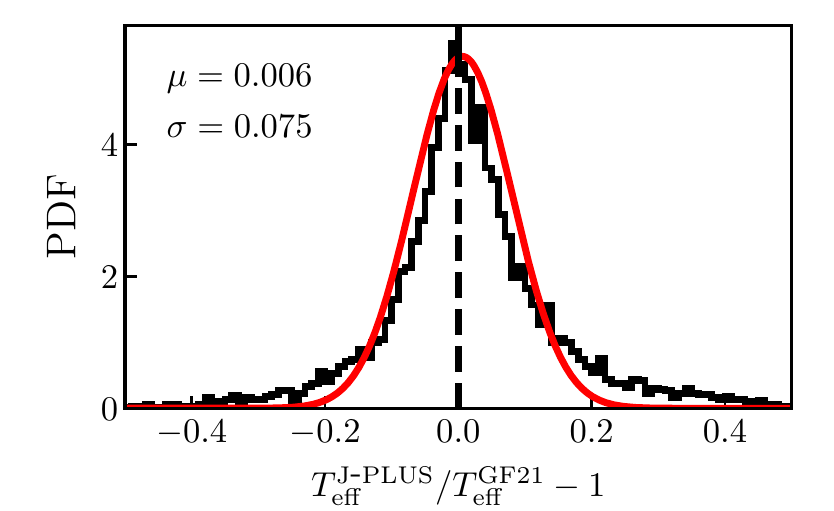}}\resizebox{0.49\hsize}{!}{\includegraphics{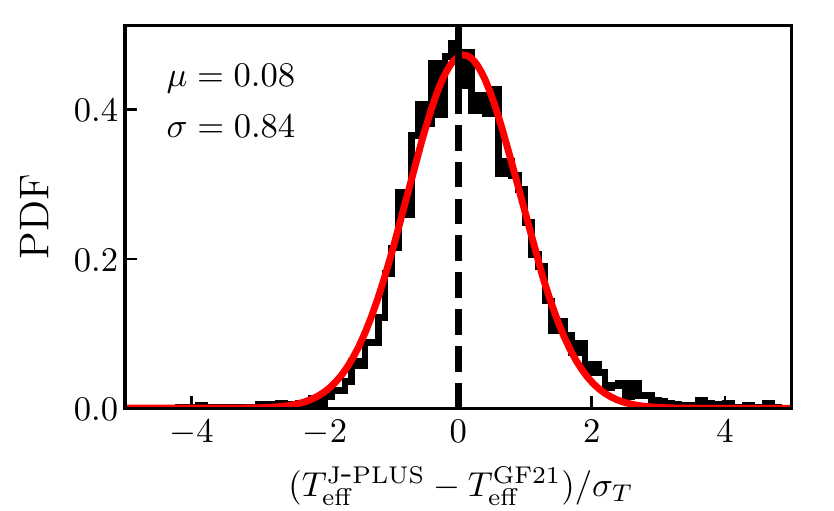}}\caption{Comparison between the effective temperature derived from J-PLUS ($T_{\rm eff}^{\rm J{\textrm -}PLUS}$) and {\it Gaia} EDR3 ($T_{\rm eff}^{\rm GF21}$) photometry. {\it Top left panel}: Individual measurements (red dots) with gray error bars. The median in ten temperature intervals is marked with the white dots. Dashed line marks the one-to-one relation. {\it Top right panel}: Relative difference between J-PLUS and {\it Gaia} temperatures as a function of {\it Gaia} temperature. White dots show the median difference in ten temperature intervals. Dashed line depicts identity. {\it Bottom left panel}: Histogram of the relative difference. {\it Bottom right panel}: Histogram of the error-normalised difference between J-PLUS and {\it Gaia} temperatures. In both {\it bottom panels}, the red line shows the best Gaussian fit to the distribution, with parameters labeled in the panel.
}
\label{fig:teffgf}
\end{figure*}

\subsection{Comparison between photometry and models}\label{sec:chidist}
Before exploring the derived white dwarf parameters, we studied the comparison between the observed photometry and the predicted fluxes from the best-fitting model. We defined the minimum $\chi^2$ of each source as
\begin{equation}
    \chi^2_{\rm min} = \sum_{j= 1}^{12}\frac{(f_j - f_j^{\rm best})^2}{\sigma^2_j} = \sum_{j= 1}^{12} (\Delta f_j)^2,
\end{equation}
where $f_j^{\rm best}$ represents the expected flux from the model with the largest probability for each analyzed white dwarf.

The histogram of the $\chi^2_{\rm min}$ for the white dwarf sample is presented in Fig.~\ref{fig:chi2dist}. We find that it is described by a $\chi^2$ distribution with 10 degrees of freedom (dof). We have 12 photometric points, implying that our modeling procedure has only two effective variables. This is a consequence of two issues: first, the surface gravity and the parallax are highly correlated and should be considered a unique effective variable. Second, the parallax prior from {\it Gaia} EDR3 (Sect.~\ref{sec:prior_px}) tightly constrains the parallax variable, and therefore the surface gravity. As a consequence, only the effective temperature and the atmospheric composition have freedom to vary in the fitting process.

The argument above was tested by repeating the Bayesian analysis but assuming a flat prior in parallax. Without the constraint from {\it Gaia} EDR3, the parallax and the surface gravity can vary in the fitting process and we expect an improvement in the minimum $\chi^2$ of the sources. We find that the distribution indeed improves (Fig.~\ref{fig:chi2dist}) and is described by a $\chi^2$ distribution with 9 dof. This means that we have three effective parameters, as anticipated.

In addition to the aggregated $\chi^2$, we can analyze the distribution of the individual passbands with respect to the best-fitting model, noted $\Delta f_j$. We expect this variable to follow a Gaussian distribution with median $\mu = 0$ and dispersion $\sigma = 0.91$, the square root of the ratio between the dof and the number of passbands. The measured distribution in $\Delta f_j$ is presented in Fig.~\ref{fig:chidist}. We find that, as desired, the median of the distributions is close to zero, with median $\mu \sim \pm 0.05$, and that the dispersion is close to expectations, with a median value of $\sigma = 0.94$ and fluctuations of only 0.1.

The median of  close to zero suggests a proper match between the photometry and the models. The color calibration of J-PLUS DR2 was performed using the locus of 639 white dwarfs \citep{clsj19jcal,clsj21zsl}, and the match found was therefore expected. The measured dispersion implies that the photometric errors are properly estimated for each passband, as they account for the observed dispersion between the photometry and the models. We conclude that the J-PLUS photometry and the estimated observational errors are reliable, providing a proper data set with which to explore the spectral evolution of the white dwarf population.

\begin{figure*}[t]
\centering
\resizebox{0.49\hsize}{!}{\includegraphics{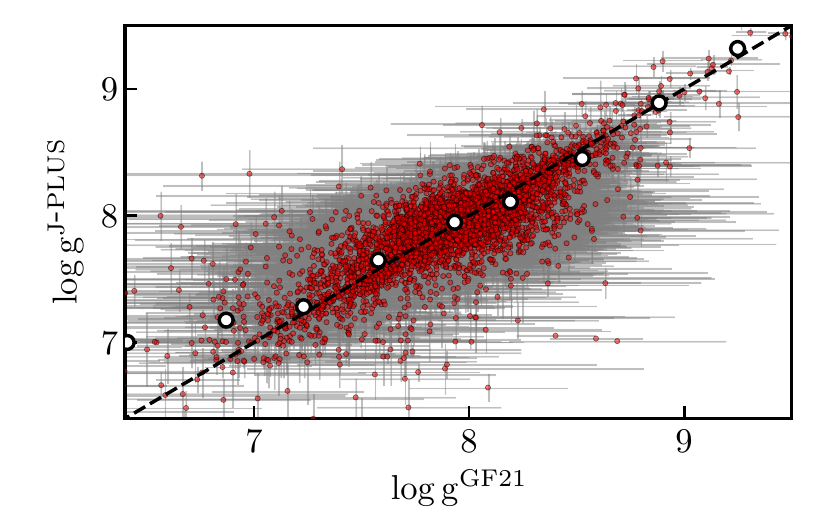}}
\resizebox{0.49\hsize}{!}{\includegraphics{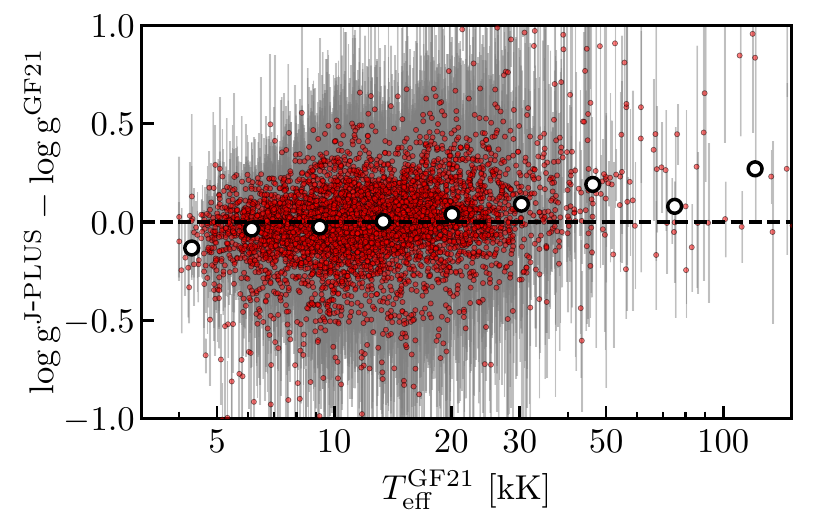}}

\resizebox{0.49\hsize}{!}{\includegraphics{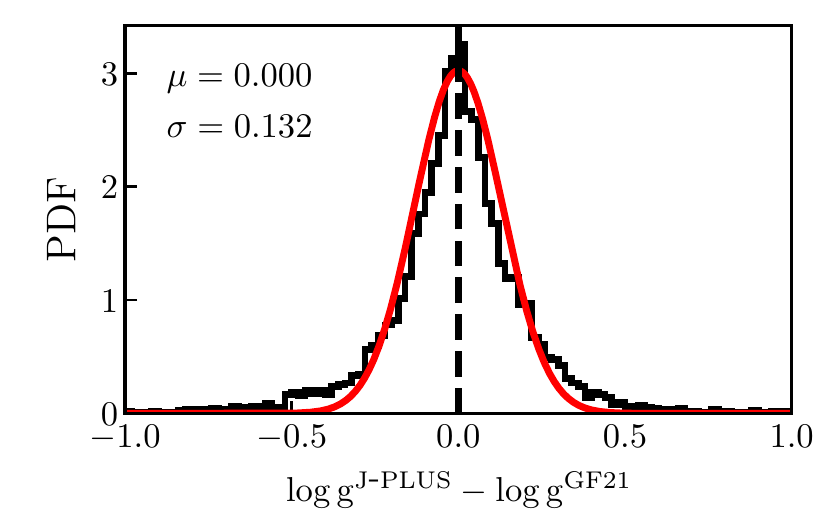}}
\resizebox{0.49\hsize}{!}{\includegraphics{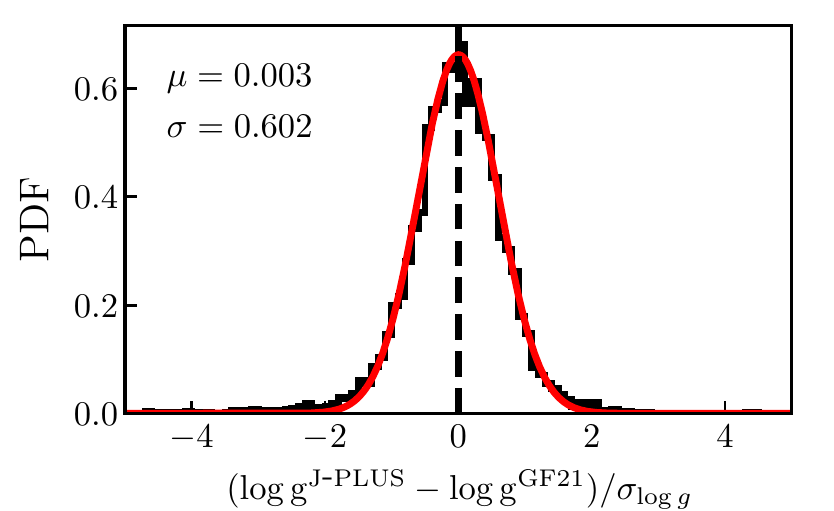}}\caption{Comparison between the surface gravity derived from J-PLUS ($\log {\rm g}^{\rm J{\textrm -}PLUS}$) and {\it Gaia} EDR3 ($\log {\rm g}^{\rm GF21}$) photometry. {\it Top left panel}: Individual measurements (red dots) with gray error bars. The median in ten gravity intervals is marked with the white dots. Dashed line marks the one-to-one relation. {\it Top right panel}: Difference between J-PLUS and {\it Gaia} surface gravity as a function of {\it Gaia} temperature. White dots mark the median difference in ten temperature intervals. Dashed line depicts identity. {\it Bottom left panel}: Histogram of the difference. {\it Bottom right panel}: Histogram of the error-normalized difference between J-PLUS and {\it Gaia} surface gravity. In both {\it bottom panels}, the red line shows the best Gaussian fit to the distribution, with parameters labeled in the panel.}
\label{fig:logggf}
\end{figure*}

\subsection{Effective temperature and surface gravity}\label{sec:tlogg}
We compare the effective temperature and the surface gravity of the $\nwd$ white dwarfs in our sample derived from J-PLUS photometry against the estimations from GF21 using {\it Gaia} EDR3 data. In the comparison, the solutions from H-dominated models in both studies were used for sources with $p_{\rm H} \geq 0.5$ and the He-dominated solutions were used otherwise. The comparison between the {\it Gaia} estimations and the values derived from spectroscopy are presented elsewhere \citep[e.g.,][GF21]{tremblay19,cukanovaite21}, and so we do not duplicate such a comparison in this work.

The comparison between J-PLUS and {\it Gaia} effective temperature scales is presented in Fig.~\ref{fig:teffgf}. We find an excellent one-to-one correlation between both measurements in the temperature range $5\,000 \lesssim T_{\rm eff} \lesssim 100\,000$ K. Moreover, the fractional difference as a function of $T_{\rm eff}$ presents no trend with temperature. The histogram of the fractional difference resembles a Lorentzian profile, with a compact core and extended wings. This is the usual distribution when measurements with different uncertainties are combined. The Gaussian approach to this distribution provides a 0.6\% difference between the two values, that is, both photometric systems provide the same effective temperature scale. The dispersion of the Gaussian distribution is $7$\%. As in Sect.~\ref{sec:chidist}, we tested the $T_{\rm eff}$ uncertainties by normalizing the difference in effective temperatures with $\sigma_{T}$, the combined error from the individual measurements. We find that the obtained distribution is well described by a Gaussian with median $\mu = 0.08$ and dispersion $\sigma = 0.84$. The median reflects the reported offset in units of the uncertainty, and the dispersion below the expected unity implies that the temperature uncertainties in both {\it Gaia} and J-PLUS could be overestimated by just $\sim 10$ \%.

The surface gravity is compared in Fig.~\ref{fig:logggf}. Both measurements are again in excellent agreement, with an apparently larger $\log {\rm g}$ in J-PLUS at $T_{\rm eff} \gtrsim 30\,000$ K. The direct comparison provides no bias and a dispersion of 0.13 dex. The error-normalized distribution resembles a Gaussian with median $\mu = 0.003$ and dispersion $\sigma = 0.60$. In this case, the dispersion is clearly below unity. We interpret this as a reflection of the correlation between the measurements, both of which used the {\it Gaia} EDR3 parallax in the analysis. The assumption of a mass--radius relation largely couples the parallax and the surface gravity parameters, which are degenerated and poorly constrained from photometry alone. Thus, the main information used to estimate $\log {\rm g}$ is shared in both studies. When accounted for in the error budget, this covariance translates to a smaller $\sigma_{\log {\rm g}}$, the combined error, and thus to a larger dispersion than in the case of having truly independent measurements. A covariance of $\rho = 0.7$ is needed to obtain a unity dispersion.

We repeated the $T_{\rm eff}$ and $\log {\rm g}$ comparison for white dwarfs with $p_{\rm H} > 0.9$ (H-dominated, $4\,011$ sources) and $p_{\rm H} < 0.1$ (He-dominated, $685$ sources). On the one hand, the results for H-dominated white dwarfs are similar to the global case, as they dominate the statistics. On the other hand, the obtained figures for the He-dominated sample are also compatible with the general case, and we only find a trend at $T_{\rm eff}^{\rm GF21} \gtrsim 20\,000$ K, with J-PLUS effective temperatures being typically lower than those based on {\it Gaia} EDR3 photometry.

The comparison between J-PLUS and {\it Gaia} values reveals satisfactory results. Unfortunately, the only net improvement is in the $T_{\rm eff}$ errors, which decrease from 10\% in {\it Gaia} to 5\% in J-PLUS, but the effective temperature and surface gravity scales are similar for both data sets. We conclude that increasing the photometric optical information from three filters in {\it Gaia} EDR3 to 12 filters in J-PLUS is not critical for $T_{\rm eff}$ and $\log {\rm g}$ estimation. As we demonstrate in the following section, the great advantage of J-PLUS with respect to {\it Gaia} EDR3 is its capability to provide the atmospheric composition of the analyzed white dwarfs. 


\section{White dwarf spectral evolution with temperature}\label{sec:fnonda_teff}
The main result of the present paper, that is, the spectral evolution in the fraction of He-dominated white dwarfs with effective temperature, is presented in Sect.~\ref{sec:fheteff}. We compare the J-PLUS results with the literature in Sect.~\ref{sec:fhe_lit}, and test the reliability of the $p_{\rm H}$ probabilities used in the estimation of the spectral evolution in Sect.~\ref{sec:pdatest}.

\subsection{Spectral evolution from J-PLUS photometry}\label{sec:fheteff}
The technical details about the reported results are presented in Sect.~\ref{sec:prior_da}. As a brief summary, the fraction of He-dominated white dwarfs ($f_{\rm He}$) is parameterized as an effective temperature function that is used as prior in the atmospheric composition and compared against the resulting $f_{\rm He}$ estimated from the posterior probabilities of the J-PLUS + {\it Gaia} white dwarf sample. The parameters of the $f_{\rm He}$ function were explored searching for self-consistency in the prior and posterior values of the He-dominated fraction.

The results from the literature (Sect.~\ref{sec:intro}) suggest that a linear function in $T_{\rm eff}$ is a proper proxy for the spectral evolution at $T_{ \rm eff} \lesssim 20\,000$ K, with a minimum fraction at the so-called DB minimum at $20\,000 \lesssim T_{\rm eff} \lesssim 45\,000$ K \citep{eisenstein06db,GB19,bedard20}. Thus, we described the spectral evolution as
\begin{eqnarray}
f_{\rm He} = b - a\times \bigg( \frac{T_{\rm eff}}{10^4\ {\rm K}} - 1 \bigg),
\end{eqnarray}
imposing a maximum value of $1$ and a minimum value of $f_{\rm He}^{\rm min}$. We defined $a$ with a negative sign to provide the evolution rate with cooling time. The three parameters that we aimed to estimate are $a$, $b$, and the minimum fraction of He-dominated white dwarfs. 

We performed the spectral analysis in the range $5\,000 < T_{\rm eff} < 30\,000$~K. As hot white dwarfs emit most of their light in the ultraviolet, optical photometry samples the Rayleigh-Jeans tail of their spectral energy distribution and is therefore weakly sensitive to their effective temperature. Our upper limit in temperature ensures a precise estimation of $T_{\rm eff}$ with J-PLUS photometry, which provides $\sigma_{\log T_{\rm eff}} \simeq 0.02$ dex at $T_{\rm eff} \lesssim 30\,000$ K. At higher temperatures, the uncertainty starts to increase, reaching $\sigma_{\log T_{\rm eff}} \simeq 0.10$ dex at $T_{\rm eff} \sim 50\,000$ K. Moreover, the maximum effective temperature in our He-dominated models is $T_{\rm eff} = 40\,000$ K, and so we also avoid undesired border effects in the solutions. The lower limit in temperature ensures that Balmer lines are visible for H-dominated white dwarfs. 

We used the full PDF in the estimation of the spectral evolution (Sect.~\ref{sec:prior_da}), and so individual sources can be spread over different temperature bins. To minimize the covariance between adjacent bins and to maximize the independent information in the calculation, we set the bin size to $\Delta \log T_{\rm eff} = 0.06 \approx 3 \times \sigma_{\log T_{\rm eff}}$. Therefore, 13 effective temperature bins were available for the spectral evolution analysis.

\begin{figure}[t]
\centering
\resizebox{\hsize}{!}{\includegraphics{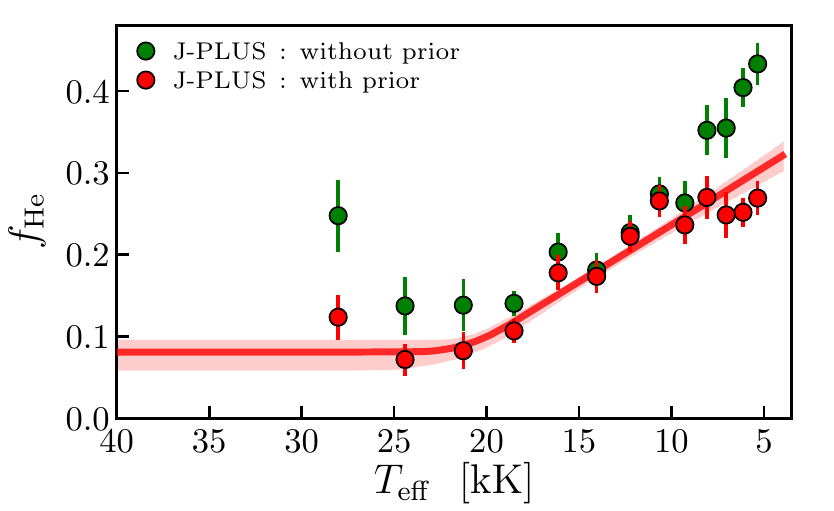}}
\caption{Fraction of He-dominated white dwarfs ($f_{\rm He}$) as a function of the effective temperature ($T_{\rm eff}$). The red solid line is the best spectral evolution prior estimated from J-PLUS photometry. The red area encloses 68\% of the solutions. The green and red dots show the values obtained from the posteriors estimated without and with the spectral evolution self-consistent prior applied, respectively.}
\label{fig:fnonda_jplus}
\end{figure}

Those sources with mass $M \leq 0.45\ M_{\odot}$ were discarded to avoid unresolved double degenerates and low-mass white dwarfs from binary evolution, leaving $4\,962$ white dwarfs. The He-dominated fraction obtained from the posteriors with and without spectral type prior are presented in Fig.~\ref{fig:fnonda_jplus} and Table~\ref{tab:fnonda}. We find that the He-dominated fraction obtained without prior has a minimum of $f_{\rm He} \simeq 0.15$ at $T_{\rm eff} \gtrsim 17\,000$ K, and then increases at lower temperatures reaching $f_{\rm He} \simeq 0.40$ at $T_{\rm eff} \sim 5\,000$ K. In other words, the J-PLUS photometry suggests a spectral evolution. However, the increase in $f_{\rm He}$ with decreasing temperature could simply be a reflection of our lower capacity to distinguish between white dwarf types. The Bayesian analysis discards such a possibility and provides a solid statistical significance to the spectral evolution.

The application of the self-consistent prior, which also provides the best measurement of the spectral evolution with temperature using J-PLUS information, yields a minimum fraction of $f_{\rm He}^{\rm min} = \fhemin \pm \fhemine$, a He-dominated fraction at $T_{\rm eff} = 10\,000$~K of $b = \bparam \pm \bparame$, and a positive slope of $a = \aparam \pm \aparame$. The parameter $a$, which reflects the rate in the spectral evolution with cooling time, is different from zero at 7$\sigma$ level.

The differences obtained in the spectral evolution with and without the spectral type prior illustrate that the capability of the J-PLUS photometry to disentangle between H- and He-dominated atmospheres degrades at higher and lower temperatures, where the spectral differences between white dwarfs with different atmospheric composition are diluted. The larger difference between the observed fraction of He-dominated white dwarfs with and without prior occurs at $T_{\rm eff} \gtrsim 17\,000$ K and $T_{\rm eff} \lesssim 9\,000$ K, where the posterior fraction decreases by 50\%. The difference is small at $9\,000 \leq T_{\rm eff} \leq 17\,000$ K, where the Balmer lines are more prominent in DAs and the contrast with respect to DBs and DCs in the J-PLUS photometry is maximum.

The final spectral evolution from J-PLUS DR2 provides a $21 \pm 3$\% increase in the fraction of He-dominated white dwarfs from $T_{\rm eff} = 20\,000$ K to $T_{\rm eff} = 5\,000$ K. We recall that the prior and the posterior fractions as a function of $T_{\rm eff}$ are self-consistent and were obtained using only J-PLUS photometric data. We demonstrate the reliability of the final $p_{\rm H}$ in Sect.~\ref{sec:pdatest}.

\begin{table} 
\caption{Spectral evolution of white dwarfs with effective temperature by PDF analysis.}
\label{tab:fnonda}
\centering 
        \begin{tabular}{c c c}
        \hline\hline\rule{0pt}{3ex} 
        $T_{\rm eff}$   &   $f_{\rm He}$ & $f_{\rm He}$  \\
              ${\rm [kK]}$       &    without prior   &   with prior      \\
        \hline\rule{0pt}{3ex}
      \!$5.34$         & $0.43 \pm 0.03$  &  $0.27 \pm 0.02$     \\ 
        $6.13$         & $0.40 \pm 0.02$  &  $0.25 \pm 0.02$     \\ 
        $7.04$         & $0.35 \pm 0.04$  &  $0.25 \pm 0.03$     \\ 
        $8.08$         & $0.35 \pm 0.03$  &  $0.27 \pm 0.03$     \\ 
        $9.28$         & $0.26 \pm 0.03$  &  $0.24 \pm 0.02$     \\ 
        $10.65$        & $0.27 \pm 0.02$  &  $0.27 \pm 0.02$     \\ 
        $12.23$        & $0.23 \pm 0.02$  &  $0.22 \pm 0.02$     \\ 
        $14.04$        & $0.18 \pm 0.02$  &  $0.17 \pm 0.02$     \\ 
        $16.13$        & $0.20 \pm 0.02$  &  $0.18 \pm 0.02$     \\ 
        $18.51$        & $0.14 \pm 0.02$  &  $0.11 \pm 0.01$     \\ 
        $21.26$        & $0.14 \pm 0.03$  &  $0.08 \pm 0.02$     \\ 
        $24.41$        & $0.14 \pm 0.03$  &  $0.07 \pm 0.02$     \\ 
        $28.02$        & $0.25 \pm 0.04$  &  $0.12 \pm 0.03$     \\
        \hline 
\end{tabular}
\end{table}

\begin{figure*}[t]
\centering
\resizebox{0.49\hsize}{!}{\includegraphics{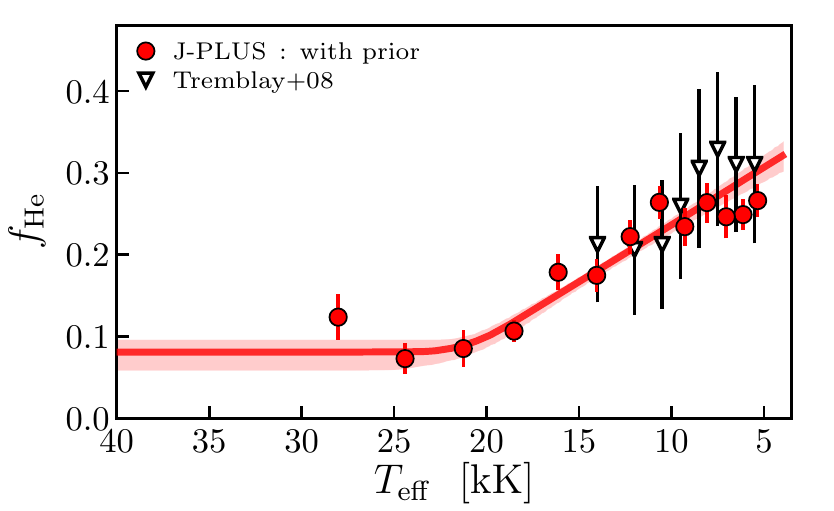}}
\resizebox{0.49\hsize}{!}{\includegraphics{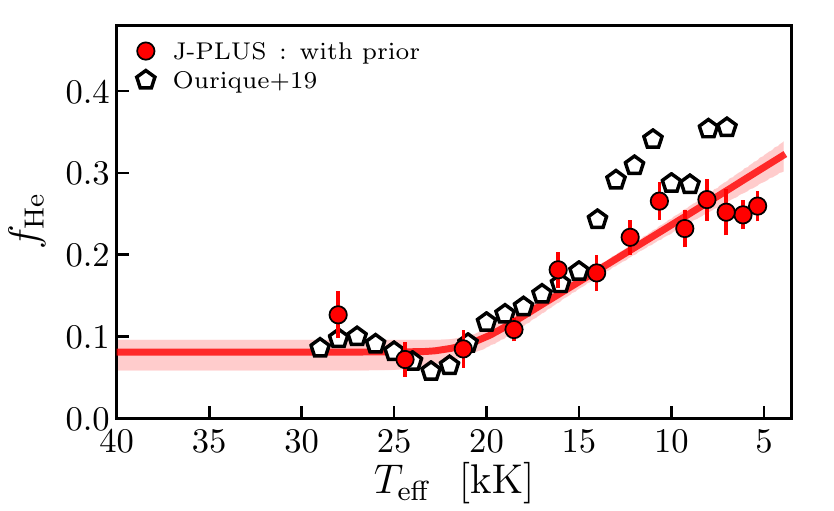}}

\resizebox{0.49\hsize}{!}{\includegraphics{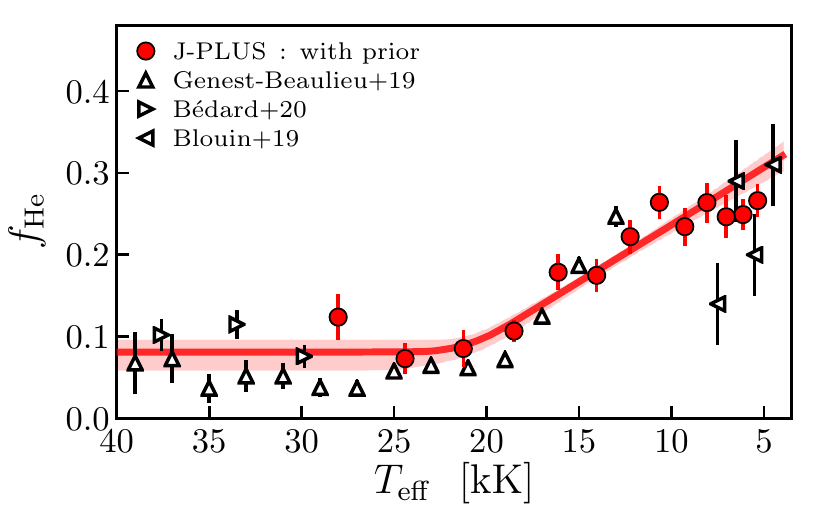}}
\resizebox{0.49\hsize}{!}{\includegraphics{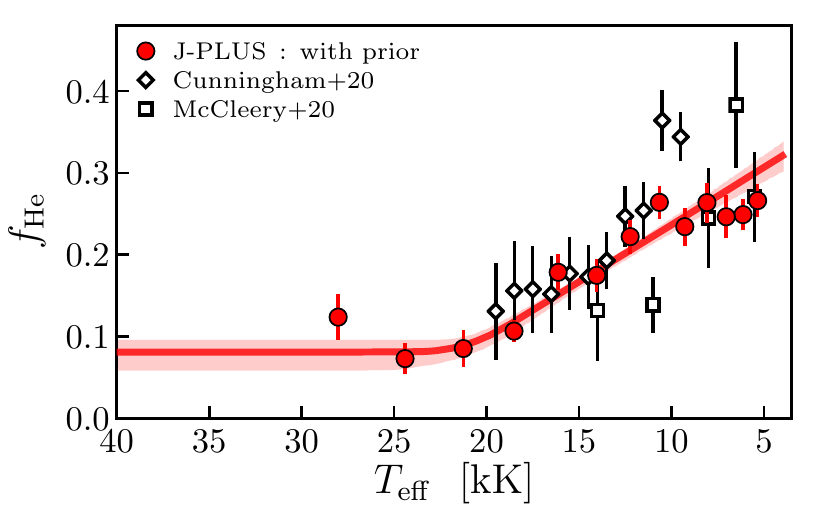}}
\caption{Fraction of He-dominated white dwarfs ($f_{\rm He}$) as a function of effective temperature ($T_{\rm eff}$) from the literature and J-PLUS. We split the comparison in four panels to improve visualization. In all panels, the red solid line is the best-fitting spectral evolution estimated from J-PLUS photometry. The red area encloses 68\% of the solutions. The red dots show the J-PLUS values obtained from the posterior estimated with the spectral evolution prior applied. The black symbols labeled in the panels show results from the literature.}
\label{fig:fnonda_lit}
\end{figure*}

\subsection{Comparison with the literature}\label{sec:fhe_lit}
We now compare our final spectral evolution with previous results in the literature, as illustrated in Fig.~\ref{fig:fnonda_lit}. There is a general agreement with the trends and values derived from spectroscopy \citep{tremblay08,GB19,ourique19,blouin19,bedard20,mccleery20} and by NUV--optical photometry \citep{cunningham20}. We recall that the J-PLUS result is based only on optical photometry.

Regarding the high-temperature end, $T_{\rm eff} \geq 20\,000$ K, the results from \citet{GB19}, \citet{ourique19}, and \citet{bedard20} suggest a lower limit of $f_{\rm He} \sim 0.05-0.10$, compatible with our derived value of $f_{\rm He}^{\rm min} = 0.08 \pm 0.02$.

At lower temperatures, previous studies found an increase in the He-dominated fraction that is compatible with the J-PLUS trend. The quantitative agreement between the J-PLUS values and the findings of \citet{tremblay08} and \citet{ourique19} 
is remarkable. We find discrepancies with \citet{ourique19} at $T_{\rm eff} \lesssim 15\,000$ K. This could be due to the nontrivial SDSS spectroscopic selection function affecting the sample used by these latter authors.

The agreement with \citet{cunningham20} is excellent over the entire temperature range, except at  $T_{\rm eff} \sim 10\,000$ K. The classification used by \citet{cunningham20} is based on the $(NUV-g)$ versus $(g-r)$ color--color diagram, where H- and He-dominated white dwarfs present different loci. The separation between these loci decreases at lower $T_{\rm eff}$, making it more difficult to classify cool white dwarfs. We note that \citet{cunningham20} do not apply a spectral type prior to their classification. This suggests that the discrepancies are just a reflection of the noisier classification at their lower temperatures and highlights the importance of spectral priors on photometric studies.

The values from \citet{mccleery20} are based on the spectroscopic follow-up of a volume-limited 40 pc sample selected from {\it Gaia} DR2 \citep{tremblay20}. The most interesting feature is the nice agreement at low temperatures, $T_{\rm eff}~<~10\,000$~K. The comparison with \citet{blouin19} at this temperature range is also satisfactory within uncertainties, but their measurements are lower than J-PLUS values in certain temperature ranges. These could be real fluctuations in the He-dominated fraction, but the smooth functional form assumed to describe the J-PLUS data is not sensitive to these possible variations, and only the general trend with $T_{\rm eff}$ can be explored. With this limitation in mind, the results from \citet{blouin19} are compatible with the J-PLUS findings.

In summary, we find good agreement with recent studies regarding the spectral evolution of white dwarfs at $T_{\rm eff} < 40\,000$~K, supporting our analysis and the unique capabilities of the J-PLUS photometric data.

\subsection{Testing the probability of having a H-dominated atmosphere}\label{sec:pdatest}
The spectral evolution presented in the previous sections is based on the H- and He-dominated posteriors estimated from J-PLUS DR2 photometry. In this section, we present three tests to check the reliability of $p_{\rm H}$: we compare the J-PLUS spectral probability with the classification from spectroscopy (Sect.~\ref{sec:spec_pda}), analyze the usual $(u-r)$ versus $(g-i)$ color--color diagram (Sect.~\ref{sec:color}), and estimate the mass distribution of H- and He-dominated white dwarfs at $d < 100$ pc (Sect.~\ref{sec:massdist}).

\subsubsection{Spectroscopic classification and significance of $p_{\rm H}$}\label{sec:spec_pda}
The atmospheric composition prior scheme developed in Sect.~\ref{sec:prior_da} can be tested by comparing $p_{\rm H}$ with the classification from spectra. We again used the spectral labels available in the GF21 catalog, estimated from SDSS DR16 spectroscopy. We only used the classification from spectra with ${\rm S/N} \geq 10$. Those white dwarfs with a dominant presence of metals in their atmosphere according to the spectroscopic classification (DZ, DZA, DZB, etc.) were included in the He-dominated class. In addition, spectral subtypes (DAB, DBA, DAZ, etc.) were assigned to their main atmospheric composition. We also restricted the sample to our temperature and mass ranges of interest, $5\,000<T_{\rm eff}<30\,000$~K and $M > 0.45$~$M_{\sun}$. We found $929$ H-dominated (DA) and $289$ He-dominated (DB/DC/DZ) white dwarfs with spectral classification.

\begin{figure}[t]
\centering
\resizebox{\hsize}{!}{\includegraphics{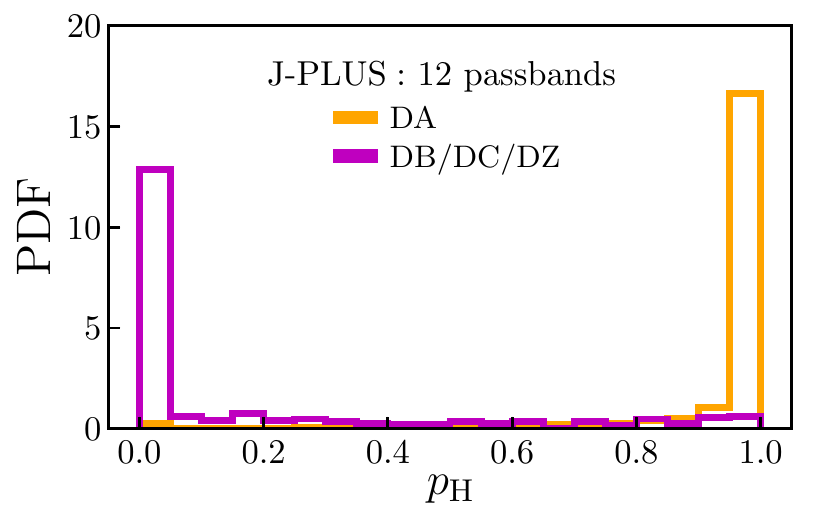}}
\caption{Normalized histogram of the $p_{\rm H}$ probability for the sample of $929$ DAs (orange) and $289$ DB/DC/DZs (purple) with spectroscopic classification.
}
\label{fig:pdahist}
\end{figure}

We first studied the $p_{\rm H}$ distribution for DA and DB/DC/DZ sources, as presented in Fig.~\ref{fig:pdahist}. We found that the DA distribution peaks at $p_{\rm H} = 1$ and the DB/DC/DZ distribution at $p_{\rm H} = 0$, as desired. Furthermore, 83\% of the DA sample have $p_{\rm H} \geq 0.95$, and 64\% of the DB/DC/DZ sample have $p_{\rm H} \leq 0.05$, demonstrating the capability of J-PLUS photometry to differentiate between different white dwarf types.

The performance of a categorical classification based on a $p_{\rm H}$ threshold can be estimated with the completeness, the purity, and other summary statistics that compare spectroscopic and photometric types. However, from a statistical point of view, we should use the measured $p_{\rm H}$ to weight the sources. In this case, all the white dwarfs are always used in the analysis and they are properly weighted with our best knowledge about their atmospheric composition. We therefore have to demonstrate that the estimated $p_{\rm H}$ is indeed the probability of having a H-dominated atmosphere.

\begin{figure}[t]
\centering
\resizebox{\hsize}{!}{\includegraphics{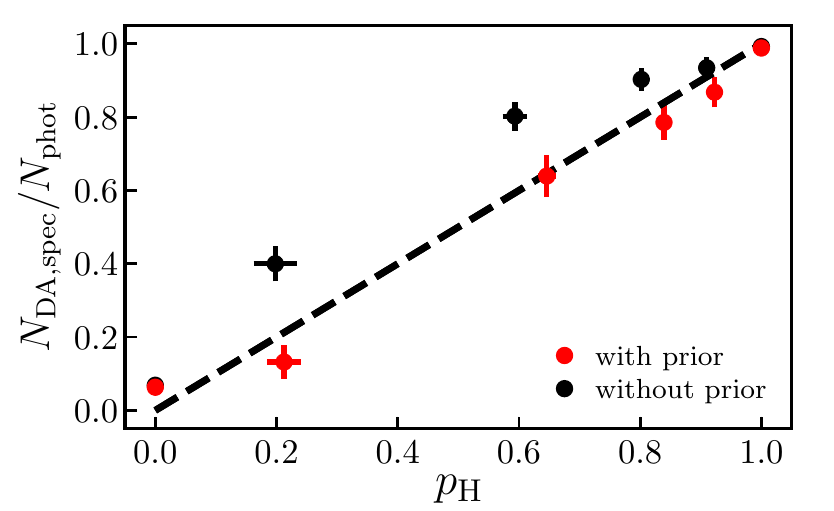}}\caption{Fraction of spectroscopic DAs as a function of $p_{\rm H}$. The black  and red dots show the results without and with the spectral type prior applied, respectively. The dashed line marks the expected one-to-one relation.}
\label{fig:pdafda}
\end{figure}

The $p_{\rm H}$ reliability can be tested by comparing the fraction of spectroscopic DAs at a given $p_{\rm H}$ range with the median $p_{\rm H}$ in that range. A properly derived $p_{\rm H}$ must produce a one-to-one relation, that is, the fraction of true DAs is proportional to $p_{\rm H}$. The obtained values are presented in Fig.~\ref{fig:pdafda} for the $p_{\rm H}$ obtained with and without atmospheric composition prior. The uncertainties were estimated from bootstrapping. We find that the relation clearly departs from the one-to-one line when the prior is not applied, with the $p_{\rm H}$ being underestimated (i.e., the fraction of true DAs is larger than predicted). The values obtained with the self-consistent prior are compatible with the desired one-to-one line. We recall that the prior was computed using J-PLUS data, and that no spectroscopic label was used in the process. This confirms the reliability of the Bayesian analysis and strengthens the spectral evolution results obtained in Sect.~\ref{sec:fheteff}.

\begin{figure*}[t]
\centering
\resizebox{0.49\hsize}{!}{\includegraphics{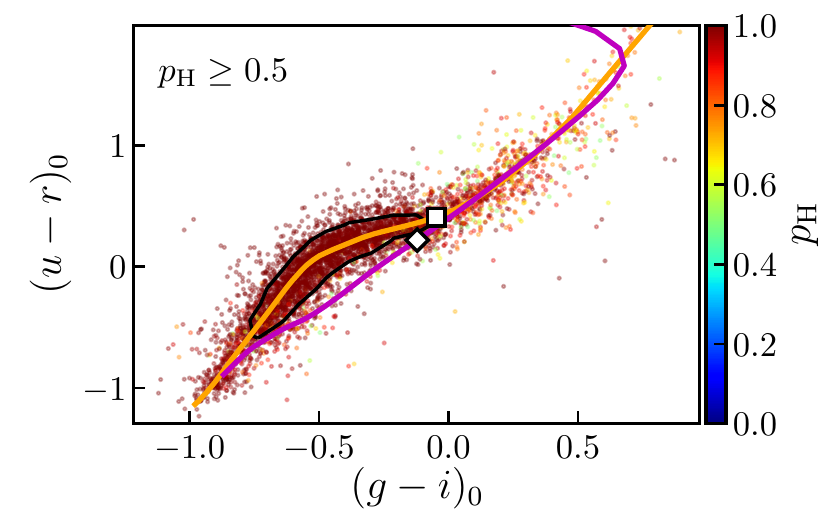}}
\resizebox{0.49\hsize}{!}{\includegraphics{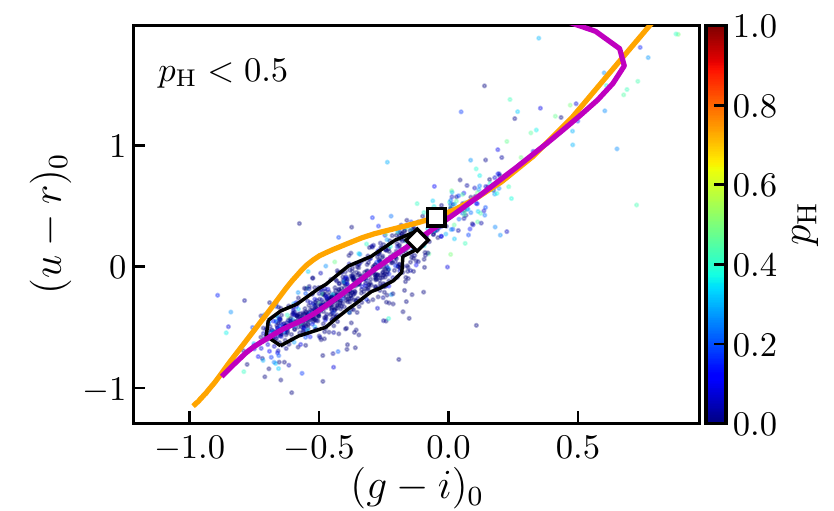}}

\resizebox{0.49\hsize}{!}{\includegraphics{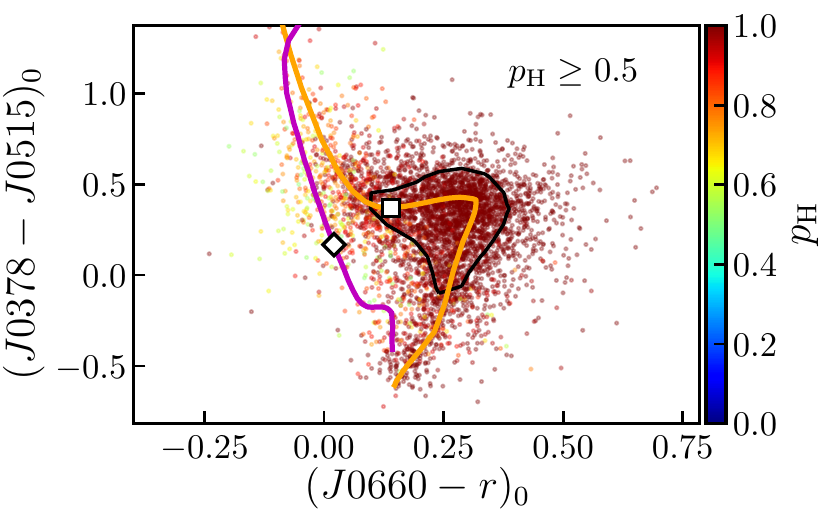}}
\resizebox{0.49\hsize}{!}{\includegraphics{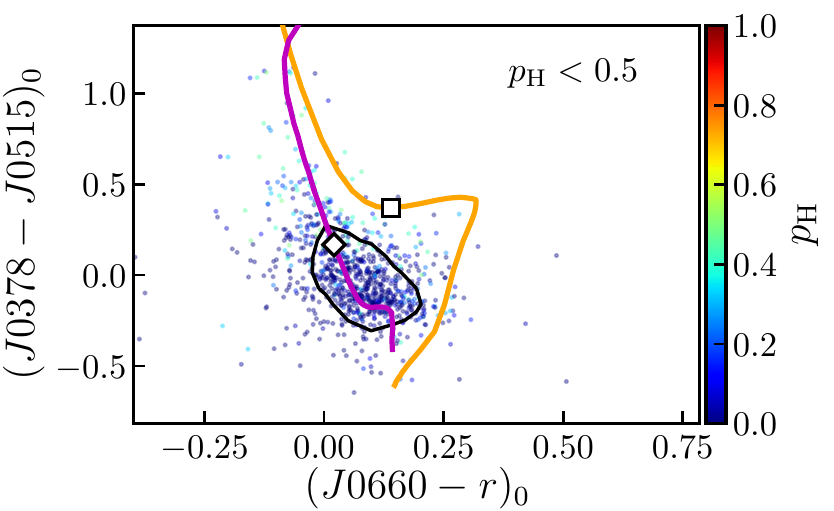}}
\caption{The $(u-r)_0$ versus $(g-i)_0$ ({\it top panels}) and $(J0378-J0515)_0$ versus $(J0660-r)_0$ ({\it bottom panels}) color--color diagrams for white dwarfs with ${\rm S/N} > 3$ in J-PLUS photometry ($5\,379$ sources). In both panels, the orange and purple lines show the theoretical loci for H-dominated and He-dominated atmospheres, respectively. The black contour enclose 50\% of the sources in each panel. The square and the diamond mark the colors for a H- and a He-dominated white dwarf, respectively, with $T_{\rm eff} = 8\,500$ K and $\log {\rm g} = 8$ dex. The color scale shows the probability of having a H-dominated atmosphere, $p_{\rm H}$. {\it Left panels}: Color--color diagrams for the $4\,470$ sources with $p_{\rm H} \geq 0.5$. {\it Right panels}: Color--color diagrams for the $909$ sources with $p_{\rm H} < 0.5$.
}
\label{fig:urgi}
\end{figure*}

As an extra test, we computed the number of H-dominated white dwarfs in the spectroscopic sample as
\begin{equation}
    N_{\rm H}^{\rm phot} = \sum_i p_{\rm H}^i,
\end{equation}
and the number of He-dominated white dwarfs as the sum of the $(1-p_{\rm H})$ probabilities. The uncertainties were again estimated by bootstrapping. We obtained $N_{\rm H}^{\rm phot} = 936 \pm 7$ and $N_{\rm He}^{\rm phot} = 282 \pm 7$ in the prior case, and $N_{\rm H}^{\rm phot} = 876 \pm 7$ and $N_{\rm He}^{\rm phot} = 342 \pm 7$ when the prior was not used. These results must be compared with the spectroscopic values $N_{\rm H}^{\rm spec} = 929$ and $N_{\rm He}^{\rm spec} = 289$. We find that the photometric and the spectroscopic numbers are compatible at 1$\sigma$ when the self-consistent prior is applied, but the discrepancies are at $7\sigma$ when the prior is neglected.

Finally, we tested the J-PLUS performance with hybrid types. There are $9$ sources classified as DAB and $26$ as DBA. We analyzed these $35$ hybrid types using the J-PLUS probabilities, obtaining $N_{\rm H}^{\rm phot} = 5 \pm 2$ and $N_{\rm He}^{\rm phot} = 30 \pm 2$ for H- and He-dominated atmospheres, respectively. The photometric values are compatible at $2\sigma$ with the spectroscopic ones. We are therefore able to obtain the main composition of the hybrid types and their presence in the white dwarf sample does not impact the classification obtained from J-PLUS data.

We conclude that the $p_{\rm H}$ derived from J-PLUS photometry is reliable and that the self-consistent prior in atmospheric composition is needed to properly recover the number of spectroscopic types. We can therefore use $p_{\rm H}$ to study the properties of H- and He-dominated white dwarfs, as illustrated in Section~\ref{sec:massdist} with the mass distribution.

\subsubsection{Color--color diagrams}\label{sec:color}
The H-dominated and He-dominated white dwarfs present two separate loci in broad-band color--color diagrams that contain the $u$ passband \citep{greenstein88}. For comparison with previous studies and to illustrate the performance of J-PLUS spectral classification, in this section we study the interstellar dust de-reddened $(u-r)_0$ versus $(g-i)_0$ color--color diagram as a function of $p_{\rm H}$ (Fig.~\ref{fig:urgi}). These plots show sources with ${\rm S/N} \geq 3$ in all J-PLUS passbands ($5\,379$ white dwarfs or 91\% of the total sample).

We find that sources with $p_{\rm H} \geq 0.5$ are clustered in the redder $(u - r)_0$ branch of the white dwarf locus at $(g-i)_0 < 0$, which corresponds to $T_{\rm eff} \gtrsim 8\,500$ K. In this temperature range, sources with $p_{\rm H} < 0.5$ are located in the bluer $(u - r)_0$ branch of the locus, as expected. At lower temperatures, the theoretical and observational colors of H- and He-dominated white dwarfs converge, making it impossible to disentangle the nature of the observed white dwarf only using broad-band optical colors. The addition of the seven J-PLUS medium-bands and the application of the self-consistent spectral type prior improves the classification below $T_{\rm eff} \sim 9\,000$~K. 

We complement this analysis with the J-PLUS color--color diagram $(J0378-J0515)_0$ versus $(J0660 - r)_0$ in the {\it bottom panels} of Fig.~\ref{fig:urgi}. This diagram includes three J-PLUS medium bands and shows part of the additional information provided by the J-PLUS filter system, which helps to better discriminate between different spectral types. On the one hand, the $(J0660 - r)$ color is sensitive to the presence of H$\alpha$ absorption. On the other hand, the combination of a J-PLUS passband with $\lambda < 4\,000\ \AA$ and the $J0515$ passband enhance the contrasts between the two types of white dwarf. An additional analysis of the white dwarf population in the J-PLUS color--color diagrams can be found in \citet{clsj19jcal}.

\begin{figure*}[t]
\centering
\resizebox{0.49\hsize}{!}{\includegraphics{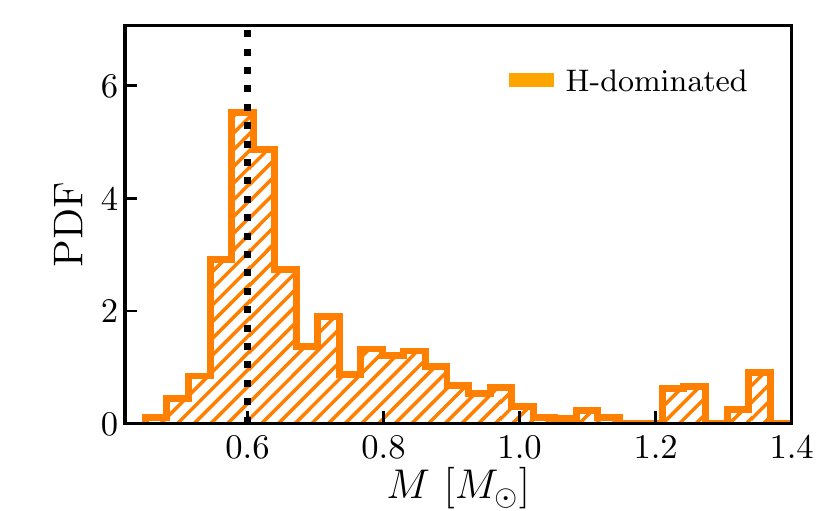}}
\resizebox{0.49\hsize}{!}{\includegraphics{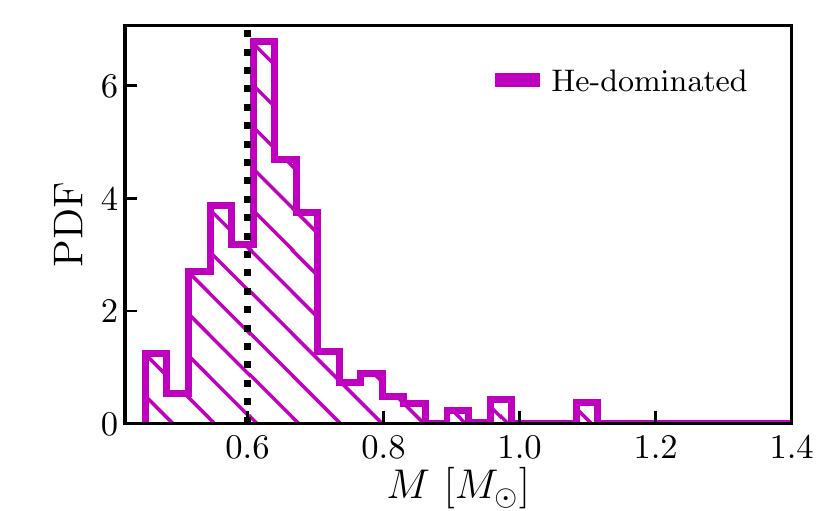}}
\caption{White dwarf mass distribution at $d \leq 100$ pc for sources with $M > 0.45\, M_{\odot}$ and $T_{\rm eff} > 6\,000$ K. {\it Left panel}: Normalized histogram weighted by $p_{\rm H}$. {\it Right panel}: Normalized histogram weighted by $(1-p_{\rm H})$. The dotted line in both panels marks a mass of $M = 0.6\, M_{\odot}$ for reference.}
\label{fig:mfwhite dwarf}
\end{figure*}

\begin{figure*}[t]
\centering
\resizebox{0.49\hsize}{!}{\includegraphics{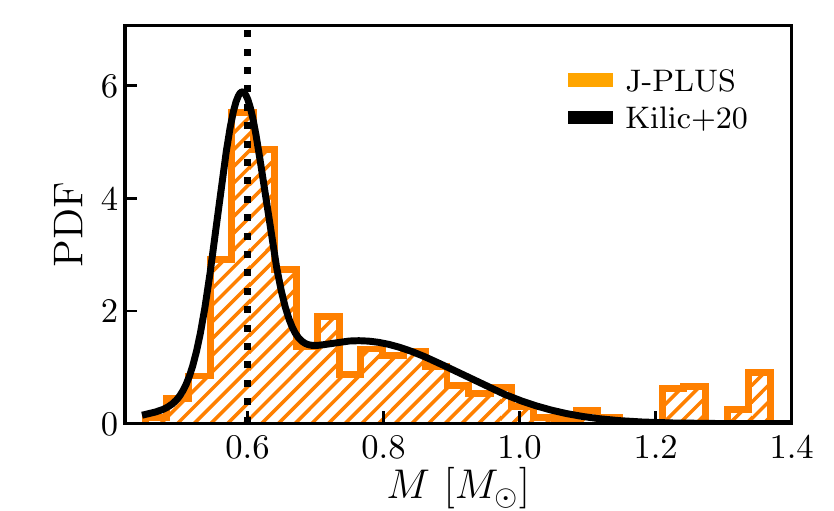}}
\resizebox{0.49\hsize}{!}{\includegraphics{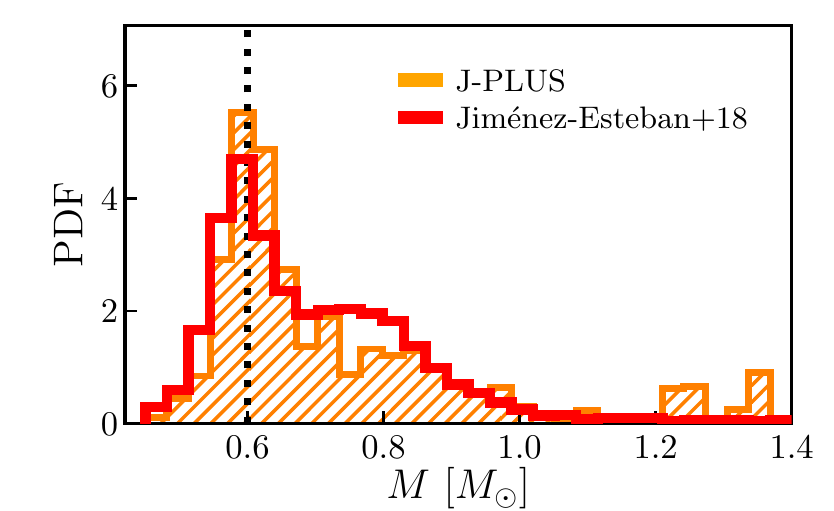}}
\caption{Comparison between the H-dominated white dwarf mass distribution at $d \leq 100$ pc, $T_{\rm eff} > 6\,000$ K, and $M > 0.45\,M_{\odot}$ from J-PLUS (dashed histograms), that of \citet[][black solid line in the {\it left panel}]{kilic20}, and that of \citet[][red solid line in the {\it right panel}]{jimenezesteban18}.
}
\label{fig:mfwhite dwarflit}
\end{figure*}

These results illustrate the performance of J-PLUS photometry in a common color--color diagram, highlighting the range of colors, $(g-i)_0 \gtrsim 0$, at which the J-PLUS data and the PDF analysis provide an advantage over broad-band photometry.  

\subsubsection{Mass distribution at $d \leq 100$ pc}\label{sec:massdist}
In this section, we estimate the stellar mass distribution of H- and He-dominated white dwarfs as a final control check for the quality of the $p_{\rm H}$ probabilities.

We restricted our sample to $d \leq 100$~pc, $T_{\rm eff} > 6\,000$~K, and $M > 0.45$~$M_{\odot}$. This selection allows us to directly compare the J-PLUS measurements with the results of \citet{jimenezesteban18} and those of \citet{kilic20}. The restricted sample in this section comprises 351 white dwarfs.

The mass distribution for H-dominated white dwarfs was estimated as the weighted histogram of the sample, where the weights were defined as
\begin{equation}
    w = p_{\rm sel} \times p_{\rm H} \times V^{-1}_{\rm eff},
\end{equation}
and the effective volume $V_{\rm eff}$ refers to the $10 \leq d \leq 100$~pc range. The weights for the He-dominated distribution were similar, but were computed with $(1 - p_{\rm H}),$  using the corresponding effective volume for He-dominated sources. The obtained distributions for H-dominated and He-dominated types, normalized to one, are presented in Fig.~\ref{fig:mfwhite dwarf}. Adding the weights, we obtain $277$ H-dominated and $84$ He-dominated white dwarfs in the restricted sample.

We find a clear peak at $M = 0.59$~M$_{\odot}$ in the mass distribution of H-dominated white dwarfs, with a high-mass tail that peaks at $M \sim 0.8$~$M_{\odot}$. This high-mass excess has been reported in several studies \citep[e.g.,][]{liebert05,limoges15,rebassa15,tremblay16,tremblay19,jimenezesteban18,kilic20}. We compare the H-dominated mass distribution with the results from \citet{jimenezesteban18} and \citet{kilic20} in Fig.~\ref{fig:mfwhite dwarflit}. We find close agreement between our results and the distribution presented by \citet{kilic20}, including the location and the amplitude of their two suggested components. These latter authors have the spectroscopic type for the sources, and we obtained similar results using our photometric classification. This result further supports our Bayesian analysis and the reliability of the $p_{\rm H}$ probabilities. 

The distribution from \citet{jimenezesteban18} presents an excess of sources at $M \sim 0.75$~$M_{\sun}$ with respect to the J-PLUS distribution. \citet{jimenezesteban18} use photometric information from the UV to the mid-infrared, but do not perform a spectral classification and assume that all the observed white dwarfs are H-dominated. They find that poor fittings are obtained for spectroscopically confirmed DBs and DCs, but this is only significant for the hotter systems, where broad bands can be use to discriminate between both types (see previous section). The contamination of cool ($T_{\rm eff} \lesssim 9\,000$~K) He-dominated white dwarfs that have typical masses of $M \sim 0.7-0.8$~$M_{\sun}$ when analyzed with pure-H models \citep{bergeron19} is a plausible explanation for the observed discrepancy.

We must mention the apparent excess of H-dominated white dwarfs with $M \gtrsim 1.2$~$M_{\odot}$ in J-PLUS. There are only six sources in the sample at this mass range, and four have $T_{\rm eff} < 10\,000$~K. Their high mass, coupled with a low effective temperature, produces a small effective volume that boosts their number density in Fig.~\ref{fig:mfwhite dwarf}. We note that our He-dominated models only reach $\log {\rm g} = 9$~dex, and so these massive white dwarfs can only be classified as H-dominated. We checked that the high mass was dictated by the parallax information, with three of our six sources being confirmed as high-mass white dwarfs by the spectroscopic follow up in \citet{kilic20}. An extra source is photometrically selected as a high-mass white dwarf candidate by \citet{kilic21}. Finally, five of our high-mass sources have $\log {\rm g} \geq 9$~dex in the \citet{jimenezesteban18} catalog.

We turn now to the mass distribution of He-dominated white dwarfs, as presented in the {\it right panel} of Fig.~\ref{fig:mfwhite dwarf}. We find a unique population located at $M = 0.62$~$M_{\odot}$. This is slightly more massive ($0.03$~$M_{\odot}$) than the primary peak for H-dominated white dwarfs, but we do not report this difference as representative. Our He-dominated models have a mixed composition with ${\rm H/He} = 10^{-5}$, as suggested by \citet{bergeron19} to reconcile the masses of DB/DCs with the masses of the DA population at $T_{\rm eff} < 11\,000$~K. We repeated the analysis with pure-He models, and the only difference that we see in the results is that the peak of the He-dominated distribution  is translated  to $M = 0.69$~$M_{\sun}$. That is, our results follow the discussion in \citet{bergeron19} and changes in the assumed He-dominated models will modify the location of the observed peak.

\begin{figure}[t]
\centering
\resizebox{\hsize}{!}{\includegraphics{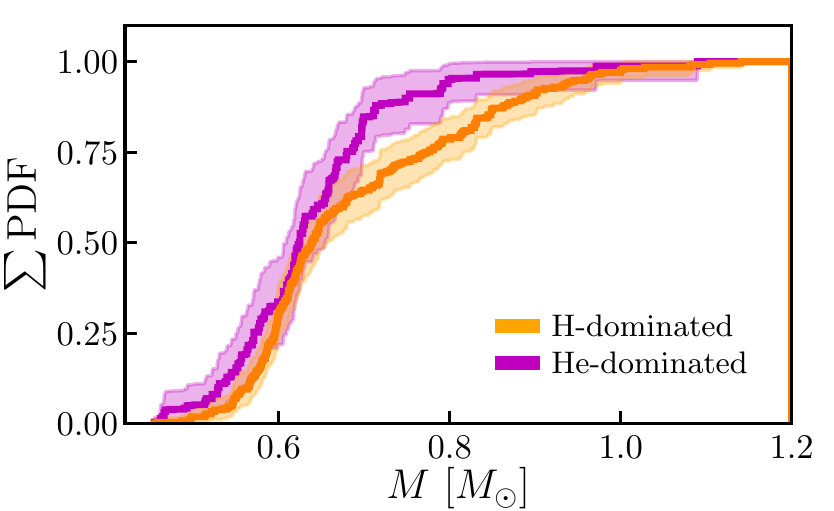}}
\caption{Cumulative mass distribution at $d \leq 100$ pc, $T_{\rm eff} > 6\,000$~K, and $0.45 < M < 1.2$~$M_{\odot}$ for H-dominated (orange histogram) and He-dominated (purple histogram) white dwarfs in J-PLUS. The colored areas show the 99\% confidence intervals in the measured distributions.}
\label{fig:mfwhite dwarfcum}
\end{figure}

Interestingly, it seems that the high-mass tail is absent in the distribution of He-dominated white dwarfs. To better illustrate this issue, the cumulative mass distributions for both types in the range $0.45 < M < 1.2$ $M_{\sun}$, including $99$\% confidence intervals estimated by bootstrapping, are presented in Fig.~\ref{fig:mfwhite dwarfcum}. The high-mass limit was imposed to avoid border effects in the comparison, as our He-dominated models were restricted to $\log {\rm g} \leq 9$~dex. There is a clear difference between both types at $M > 0.65$~$M_{\sun}$, where the excess of H-dominated white dwarfs with respect to the He-dominated distribution seems significant at a level of more than 99\%. We performed a two-sample Kolmogorov-Smirnov test, finding that the maximum difference between the cumulative curves, $D = 0.22$, corresponds to a 0.4\% probability that both distributions were extracted from the same parent population. That is equivalent to a $3\sigma$ significance in the observed difference. The lack of a high-mass tail in the He-dominated distribution has previously been reported using spectroscopic classifications \citep[e.g.,][]{bergeron01,tremblay19}, and we reproduce this result here using only optical photometric data.

We conclude that the statistical type classification from J-PLUS photometry is able to provide reliable mass distributions for H- and He-dominated white dwarfs.


\section{Summary and conclusions}\label{sec:conclusions}
We analyzed a sample of $\nwd$ white dwarfs with $r \leq 19.5$ mag in common between the {\it Gaia} EDR3 catalog from \citet{GF21} and J-PLUS DR2. We estimated the effective temperature, surface gravity, parallax, and atmospheric composition (H-dominated or He-dominated) of the sources with a Bayesian analysis. We used the parallax from {\it Gaia} EDR3 as prior, and derived a self-consistent prior for the atmospheric composition as a function of $T_{\rm eff}$ using J-PLUS photometric data alone. A way to access to the derived parameters is described in Appendix~\ref{app:data}.

We find that the fraction of He-dominated white dwarfs ($f_{\rm He}$) increases by $21 \pm 3$\% from $T_{\rm eff} = 20\,000$~K to $T_{\rm eff} = 5\,000$~K. We describe the fraction of He-dominated white dwarfs with a linear function of the effective temperature at $5\,000 \leq T_{\rm eff} \leq 30\,000\ {\rm K}$. We find $f_{\rm He} = \bparam \pm \bparame$ at $T_{\rm eff} = 10\,000$~K, a change rate along the cooling sequence of $\aparam \pm \aparame$ per $10$~kK, and a minimum He-dominated fraction of $f_{\rm He}^{\rm min} = \fhemin \pm \fhemine$ at the high-temperature end. The derived values of the He-dominated fraction are in agreement with previous results in the literature, where the observed spectral evolution is interpreted as the effect of convective mixing and convective dilution.

We tested the estimated probabilities of being H-dominated by comparison with the classification from spectroscopy. We find that the derived $p_{\rm H}$ provides the true probability of being H-dominated, so it can be used to obtain reliable probability-weighted distributions of the white dwarf population. We highlighted the last point by estimating the mass distribution at $d \leq 100$ pc and $T_{\rm eff} > 6\,000$ K for H- and He-dominated white dwarfs. Our findings for the H-dominated distribution resemble those of previous work, with a dominant $M = 0.59$~$M_{\odot}$ peak and the presence of a high-mass tail at $M \sim 0.8$~$M_{\odot}$. This high-mass excess is absent in the He-dominated distribution, which presents a single peak at $M \simeq 0.6$~$M_{\odot}$.

This work also provides hints about the capabilities of low-spectral-resolution data ($R \sim 50$) in the study of the white dwarf population. The future spectro-photometry from {\it Gaia} DR3 and the photo-spectra from the Javalambre Physics of the accelerating Universe Astrophysical Survey (J-PAS, 56 narrow-bands of 14 nm width in the optical over thousands of square degrees up to $m \sim 22.5$ mag; \citealt{jpas, minijpas}) coupled with the massive spectroscopic follow up from the William Herschel Telescope Enhanced Area Velocity Explorer (WEAVE; \citealt{weave}) will expand our knowledge of white dwarf spectral evolution with large and homogeneous data sets.


\begin{acknowledgements}
We dedicate this paper to the memory of our six IAC colleagues and friends who met with a fatal accident in Piedra de los Cochinos, Tenerife, in February 2007, with  special thanks to Maurizio Panniello, whose teachings of \texttt{python} were so important for this paper.

We thank the discussions with the members of the J-PLUS collaboration, especially to A.~J.~Dom\'{\i}nguez-Fern\'andez. We thank the anonymous referee for useful comments and suggestions.

Based on observations made with the JAST80 telescope at the Observatorio Astrof\'{\i}sico de Javalambre (OAJ), in Teruel, owned, managed, and operated by the Centro de Estudios de F\'{\i}sica del  Cosmos de Arag\'on. We acknowledge the OAJ Data Processing and Archiving Unit (UPAD, \citealt{upad}) for reducing and calibrating the OAJ data used in this work.

Funding for the J-PLUS Project has been provided by the Governments of Spain and Arag\'on through the Fondo de Inversiones de Teruel; the Aragonese Government through the Reseach Groups E96, E103, and E16\_17R; the Spanish Ministry of Science, Innovation and Universities (MCIU/AEI/FEDER, UE) with grants PGC2018-097585-B-C21 and PGC2018-097585-B-C22; the Spanish Ministry of Economy and Competitiveness (MINECO) under AYA2015-66211-C2-1-P, AYA2015-66211-C2-2, AYA2012-30789, and ICTS-2009-14; and European FEDER funding (FCDD10-4E-867, FCDD13-4E-2685). The Brazilian agencies FINEP, FAPESP, and the National Observatory of Brazil have also contributed to this project.

P.~-E.~T. has received funding from the European Research Council under the European Union's Horizon 2020 research and innovation programme n. 677706 (white dwarf3D). 

J.~M.~C. acknowledge financial support by the Spanish Ministry of Science, Innovation and University (MICIU/FEDER, UE) through grant RTI2018-095076-B-C21, and the Institute of Cosmos Sciences University of Barcelona (ICCUB, Unidad de Excelencia ’Mar\'{\i}a de Maeztu’) through grant CEX2019-000918-M.

J.~V. acknowledges the technical members of the UPAD for their invaluable work: Juan Castillo, Javier Hern\'andez, \'Angel L\'opez, Alberto Moreno, and David Muniesa.

B.~T.~G. was supported by STFC grantST/T000406/1 and by a Leverhulme
Research Fellowship.

F.~J.~E. acknowledges financial support from the Spanish MINECO/FEDER
through the grant AYA2017-84089 and MDM-2017-0737 at Centro de
Astrobiolog\'{\i}a (CSIC-INTA), Unidad de Excelencia Mar\'{\i}a de Maeztu, and
from the European Union’s Horizon 2020 research and innovation programme
under Grant Agreement no. 824064 through the ESCAPE - The European
Science Cluster of Astronomy \& Particle Physics ESFRI Research
Infrastructures project.

R.~A.~D. acknowledges support from the CNPq through BP grant 308105/2018-4.

This work has made use of data from the European Space Agency (ESA) mission
{\it Gaia} (\url{https://www.cosmos.esa.int/gaia}), processed by the {\it Gaia} Data Processing and Analysis Consortium (DPAC, \url{https://www.cosmos.esa.int/web/gaia/dpac/consortium}). Funding for the DPAC has been provided by national institutions, in particular the institutions participating in the {\it Gaia} Multilateral Agreement.

We also used the data from the SVO archive of White Dwarfs from {\it Gaia} (\url{http://svo2.cab.inta-csic.es/vocats/v2/wdw/}) at CAB (INTA-CSIC).
Funding for SDSS-III has been provided by the Alfred P. Sloan Foundation, the Participating Institutions, the National Science Foundation, and the U.S. Department of Energy Office of Science. The SDSS-III web site is \url{http://www.sdss3.org/}.

SDSS-III is managed by the Astrophysical Research Consortium for the Participating Institutions of the SDSS-III Collaboration including the University of Arizona, the Brazilian Participation Group, Brookhaven National Laboratory, Carnegie Mellon University, University of Florida, the French Participation Group, the German Participation Group, Harvard University, the Instituto de Astrofisica de Canarias, the Michigan State/Notre Dame/JINA Participation Group, Johns Hopkins University, Lawrence Berkeley National Laboratory, Max Planck Institute for Astrophysics, Max Planck Institute for Extraterrestrial Physics, New Mexico State University, New York University, Ohio State University, Pennsylvania State University, University of Portsmouth, Princeton University, the Spanish Participation Group, University of Tokyo, University of Utah, Vanderbilt University, University of Virginia, University of Washington, and Yale University.

Funding for the Sloan Digital Sky Survey IV has been provided by the  Alfred P. Sloan Foundation, the U.S. Department of Energy Office of Science, and the Participating  Institutions. SDSS-IV acknowledges support and resources from the Center for High Performance Computing  at the  University of Utah. The SDSS  website is \url{www.sdss.org}. SDSS-IV is managed by the Astrophysical Research Consortium for the Participating Institutions of the SDSS Collaboration including the Brazilian Participation Group, the Carnegie Institution for Science, Carnegie Mellon University, Center for Astrophysics | Harvard \& Smithsonian, the Chilean Participation Group, the French Participation Group, Instituto de Astrof\'isica de Canarias, The Johns Hopkins University, Kavli Institute for the Physics and Mathematics of the Universe (IPMU) / University of Tokyo, the Korean Participation Group, Lawrence Berkeley National Laboratory, Leibniz Institut f\"ur Astrophysik Potsdam (AIP),  Max-Planck-Institut f\"ur Astronomie (MPIA Heidelberg), Max-Planck-Institut f\"ur Astrophysik (MPA Garching), Max-Planck-Institut f\"ur Extraterrestrische Physik (MPE), National Astronomical Observatories of China, New Mexico State University, New York University, University of Notre Dame, Observat\'ario Nacional / MCTI, The Ohio State University, Pennsylvania State University, Shanghai Astronomical Observatory, United Kingdom Participation Group, Universidad Nacional Aut\'onoma de M\'exico, University of Arizona, University of Colorado Boulder, University of Oxford, University of Portsmouth, University of Utah, University of Virginia, University of Washington, University of Wisconsin, Vanderbilt University, and Yale University.

This research made use of \texttt{Astropy}, a community-developed core \texttt{Python} package for Astronomy \citep{astropy}, and \texttt{Matplotlib}, a 2D graphics package used for \texttt{Python} for publication-quality image generation across user interfaces and operating systems \citep{pylab}.
\end{acknowledgements}

\bibliographystyle{aa}
\bibliography{biblio}

\newpage

\begin{appendix}

\section{Access to the white dwarf atmospheric parameters and composition}\label{app:data}
\begin{table*}[t] 
\caption{White dwarf catalog of atmospheric parameters and composition.}
\label{tab:catalog}
\centering 
        \begin{tabular}{c c l}
        \hline\hline\rule{0pt}{3ex} 
        Heading     & Units  &  Description   \\
        \hline\rule{0pt}{3ex}
      \!TILE\_ID     &  $\cdots$       & Identifier of the J-PLUS Tile image in the $r$ band where the object was detected. \\ 
        NUMBER       &  $\cdots$       & Number identifier assigned by \texttt{SExtractor} for the object in the $r$-band image. \\
        RAdeg        &  deg            & Right ascension (J2000). \\
        DEdeg        &  deg            & Declination (J2000). \\
        rmag         &  mag            & De-reddened total $r-$band apparent magnitude, noted $\hat{r}$.     \\ 
        e\_rmag      &  mag            & Uncertainty in $\hat{r}$.\\ 
        p\_sel        &  $\cdots$       & Selection probability for $\hat{r} \leq 19.5$ mag and $1 \leq \varpi \leq 100$ mas.\\ 
        p\_H          &  $\cdots$       & Probability of having a H-dominated atmosphere.\\ 
        Teff\_H      &  K              & Effective temperature ($T_{\rm eff}$) for a H-dominated atmosphere. \\ 
        e\_Teff\_H   &  K              & Uncertainty in Teff\_H.\\
        logg\_H      &  dex            & Decimal logarithm of the surface gravity ($\log {\rm g}$) for a H-dominated atmosphere. \\ 
        e\_logg\_H   &  dex            & Uncertainty in logg\_H.\\
        plx\_H       &  mas            & Parallax for a H-dominated atmosphere. \\ 
        e\_plx\_H    &  mas            & Uncertainty in plx\_H.\\
        mass\_H      &  $M_{\odot}$    & Mass for a H-dominated atmosphere. \\ 
        e\_mass\_H   &  $M_{\odot}$    & Uncertainty in mass\_H.\\ 
        V\_H      &  ${\rm pc}^3$      & Effective volume ($V$) at $10 \leq d \leq 1000$ pc for a H-dominated atmosphere. \\ 
        Teff\_He     &  K              & Effective temperature for a He-dominated atmosphere. \\ 
        e\_Teff\_He  &  K              & Uncertainty in Teff\_He.\\
        logg\_He     &  dex            & Decimal logarithm of the surface gravity for a He-dominated atmosphere. \\ 
        e\_logg\_He  &  dex            & Uncertainty in logg\_He.\\
        plx\_He      &  mas            & Parallax for a He-dominated atmosphere. \\ 
        e\_plx\_He   &  mas            & Uncertainty in plx\_He.\\
        mass\_He     &  $M_{\odot}$    & Mass for a He-dominated atmosphere. \\ 
        e\_mass\_He  &  $M_{\odot}$    & Uncertainty in mass\_He.\\ 
        V\_He     &  ${\rm pc}^3$      & Effective volume at $10 \leq d \leq 1000$ pc for a He-dominated atmosphere. \\ 
        \hline 
\end{tabular}
\end{table*}

In this Appendix, we provide detailed information about access to the atmospheric parameters and composition of the $\nwd$ white dwarfs analyzed in the present paper. The gathered information is accessible in the J-PLUS database\footnote{\url{http://archive.cefca.es/catalogues/jplus-dr2/tap_async.html}} at the table \texttt{jplus.WhiteDwarf}, which is also accessible through virtual observatory table access protocol (TAP) service\footnote{\url{https://archive.cefca.es/catalogues/vo/tap/jplus-dr2}}, and at the CDS. The description of the columns is provided in Table.~\ref{tab:catalog}. 

Here, we present some ADQL queries as an example for the user. The full catalog is retrieved as
\begin{verbatim}
    SELECT *
    FROM jplus.WhiteDwarf
\end{verbatim}

The information of the six massive white dwarfs discussed in Sect.~\ref{sec:massdist} is obtained with
\begin{verbatim}
    SELECT *
    FROM jplus.WhiteDwarf 
    WHERE mass_H > 1.2
    AND Teff_H > 6000 AND Teff_H < 30000
    AND plx_H >= 10
\end{verbatim}

The relevant information used to estimate the mass distribution of He-dominated atmosphere white dwarfs in Sect.~\ref{sec:massdist} is downloaded with
\begin{verbatim}
    SELECT p_sel,p_H,mass_He,V_He 
    FROM jplus.WhiteDwarf 
    WHERE p_H < 1 AND mass_He > 0.45 
    AND Teff_He > 6000 AND Teff_He < 30000
    AND plx_He >= 10
\end{verbatim}

The information in the catalog is also accessible with the J-PLUS explorer of individual objects, including a graphical view of the best-fitting solution as in Fig.~\ref{fig:examples}.

\end{appendix}

\end{document}